\documentclass[usenatbib]{mnras}
\usepackage[utf8]{inputenc}
\usepackage{amsmath}
\usepackage{amssymb}
\usepackage{graphicx}
\usepackage[T1]{fontenc}
\usepackage{ae,aecompl}
\usepackage{subfig}

\title[Yule-Simpson's paradox in Galactic Archaeology]
  {Yule-Simpson's paradox in Galactic Archaeology}

\author[I.~Minchev et al.]
{I.~Minchev,$^{1}$
G.~Matijevic,$^{1}$
D.~W.~Hogg,$^{2,3,4,5,}$
G.~Guiglion,$^{1}$
M.~Steinmetz,$^{1}$
\newauthor
F.~Anders,$^{1}$
C.~Chiappini,$^{1}$
M.~Martig,$^6$
A.~Queiroz,$^{1}$
C.~Scannapieco$^{7}$ 
\\
$^{1}$Leibniz-Institut f\"ur  Astrophysik Potsdam (AIP), An der Sternwarte 16, 14482 Potsdam, Germany\\
$^{2}${Center for Cosmology and Particle Physics, Department of Physics, New York University, 726 Broadway, New York, NY 10003, USA}\\
$^{3}${Center for Data Science, New York University, 60 Fifth Ave, New York, NY 10011, USA}\\
$^{4}${Max-Planck-Institut f\"ur Astronomie, K\"onigstuhl 17, D-69117 Heidelberg}\\
$^{5}${Flatiron Institute, Simons Foundation, 162 Fifth Ave, New York, NY 10010, USA}\\
$^6$Astrophysics Research Institute, Liverpool John Moores University, 146 Brownlow Hill, Liverpool L3 5RF, UK\\
$^7$Universidad de Buenos Aires, Facultad de Ciencias Exactas y Naturales, Departamento de F\'sica. Buenos Aires, Argentina
}

\pagerange{\pageref{firstpage}--\pageref{lastpage}} \pubyear{2019}

\pubyear{2019}

\begin{document}
\label{firstpage}
\pagerange{\pageref{firstpage}--\pageref{lastpage}}
\maketitle

\begin{abstract}

Simpson's paradox, or Yule-Simpson effect, arises when a trend appears in different subsets of data but disappears or reverses when these subsets are combined. We describe here seven cases of this phenomenon for chemo-kinematical relations believed to constrain the Milky Way disk formation and evolution. We show that interpreting trends in relations, such as the radial and vertical chemical abundance gradients, the age-metallicity relation, and the metallicity-rotational velocity relation (MVR), can lead to conflicting conclusions about the Galaxy past if analyses marginalize over stellar age and/or birth radius. It is demonstrated that the MVR in RAVE giants is consistent with being always strongly negative, when narrow bins of [Mg/Fe] are considered. This is directly related to the negative radial metallicity gradients of stars grouped by common age (mono-age populations) due to the inside-out disk formation. The effect of the asymmetric drift can then give rise to a positive MVR trend in high-[$\alpha$/Fe] stars, with a slope dependent on a given survey's selection function and observational uncertainties. We also study the variation of lithium abundance, A(Li), with [Fe/H] of AMBRE:HARPS dwarfs. A strong reversal in the positive A(Li)-[Fe/H] trend of the total sample is found for mono-age populations, flattening for younger groups of stars. Dissecting by birth radius shows strengthening in the positive A(Li)-[Fe/H] trend, shifting to higher [Fe/H] with decreasing birth radius; these observational results suggest new constraints on chemical evolution models.
This work highlights the necessity for precise age estimates for large stellar samples covering wide spatial regions. 
\end{abstract}


\section{Introduction}
\label{sec:intro}

Galactic Archaeologists use the observed stellar kinematics, chemical abundances, and derived ages as fossil records to recover the chemo-dynamical history of the Milky Way \citep{freeman02, bland-hawthorn10, rix13, binney13, minchev16a, hogg16}. Age information is by far the hardest one to obtain, yet extremely important to break degeneracies among different chemical evolution models \citep{miglio17}.

To infer the Galactic disk history, we rely on observed correlations, such as  the mean stellar metallicity\footnote{As commonly done in the literature, throughout this work we use 'metallicity' to mean the iron fraction [Fe/H], not [M/H] - the overall metallicity of a star, defined as the total amount of elements heavier than Helium.}, [Fe/H], as a function of Galactic disk radius. The negative metallicity gradient observed today suggests that the disk has formed inside out \citep{matteucci89}. The observational situation is improving rapidly: With the expansion of the disk coverage around the Sun thanks to a number of Galactic surveys, it has become possible in recent years to identify a number of new chemo-kinematical trends. 

For example, an inversion of the radial metallicity gradient from negative to positive (and in [$\alpha$/Fe] gradient from positive to negative) for samples at increasingly larger distance from the disk midplane, $|z|$, has been observed in most large Milky Way surveys (e.g., SEGUE - \citealt{cheng12a}, RAVE - \citealt{boeche13b}, APOGEE - \citealt{anders14}, Gaia-ESO - \citealt{recio-blanco14}, LAMOST - \citealt{wang19}), explained by \cite{mcm14} (see also \citealt{rahimi14, miranda16, ma17, schonrich17}) as the effect of inside-out disk formation and disk flaring (increasing disk thickness with Galactocentric radius) of mono-age populations (Fig.~10 in \citealt{mcm14}). It was shown that gradient inversion can result from the variation in the stellar age distribution with radius: at large $|z|$ stars that are old, metal-poor, and  [$\alpha$/Fe]-rich dominate in the inner disk (due to the inside-out formation), while younger, more metal-rich, and more [$\alpha$/Fe]-poor stars are more abundant in the outer disk (present at large $|z|$ thanks to disk flaring). This naturally results in a positive metallicity (and a negative [$\alpha$/Fe]) gradient at large $|z|$, although for groups of stars of common age (mono-age populations) the gradient is always negative.

Another example of a trend that can be linked to the Galaxy's past is disk flaring (e.g., \citealt{kazantzidis08, bovy16, kawata17, wang18, bland-hawthorn16}). Disk flaring can be related to the dynamical effect of merging satellites, e.g., the Sagittarius dwarf \citep{laporte18a, thomas19}. Using two suites of simulations of galactic disk formation in the cosmological context, \cite{minchev15} showed that flaring is always present for mono-age populations. When considering the total stellar population, however, it was found that the disk flaring was lost or at least significantly decreased. This was explained by a combination of the nested flares of mono-age populations, where younger groups of stars flare and dominate in density at progressively larger radii as a result of the inside-out disk formation.

As a final example, \cite{mcm13} (hereafter MCM13) found that in the solar vicinity resulting from their chemo-dynamical model the variation of [Fe/H] with vertical distance from the disk midplane, $|z|$, showed a strong negative gradient ($\sim-0.3$~dex/kpc) but almost flat relations for mono-age populations (see their Fig.~9, top). This phenomenon could be understood by taking into account that the fraction of metal-poor old stars increases with $|z|$, thus giving rise to a negative gradient for the total population (see results with SEGUE G-dwarfs -- \citealt{schlesinger12}, RAVE -- \citealt{boeche13b}, and LAMOST data -- \citealt{wang19}).

What all of the above examples have in common is that a relation resulting from the total sample is weakened, erased, or even inverted, compared to the trends in subsets grouped by considering a third variable (in this case, age).

This phenomenon is known as Simpson's, or Yule-Simpson's, paradox (hereafter, YSP), in classical statistics (e.g., \citealt{yule02, simpson51,smith12, thompson06}). It has mostly been described in social sciences, medicine, biology, and baseball batting averages, but also in quantum mechanics \citep{li13, selvitella17}. YSP arises when both the independent and dependent variables of a given relation depend on yet a third variable, often referred to as a "lurking" or "confounding" variable (in the above cases  - age). A lurking variable is one for which data are unavailable, but it nevertheless has influence on other variables in the study. See, for example, \cite{brase16} for more information on the topic.

Five cases of the YSP, although not identified as such, occurring in the field of Galactic Archaeology have been described in several works in the last five years \citep{mcm13,mcm14,minchev14,minchev15,minchev18}. We here elaborate on those and discuss two new cases. We will refer to a "weak" case of Simpson's paradox when a relation is simply flattened or weakened and a "strong" case when it is completely inverted.

In \S\S\ref{sec:grad}-\ref{sec:mvr} we make use of the Milky Way chemo-dynamical model by MCM13, who used a hybrid technique to combine a classical semi-analytical chemical evolution model \citep{chiappini09a} with a simulation in the cosmological context \citep{martig12}. The model has been shown to comply with a range of observational constraints, as well as predict new ones (for a summary, see \citealt{minchev16a}). In \S\ref{sec:grad} and \S\ref{sec:flaring} we present YSP cases for global Galactic disk chemo-kinematical trends. In \S\S\ref{sec:rb}-\ref{sec:li} we will be concerned with relations of samples confined to the solar neighborhood.

\begin{figure}
\centering
\includegraphics[width=0.9\linewidth, angle=0]{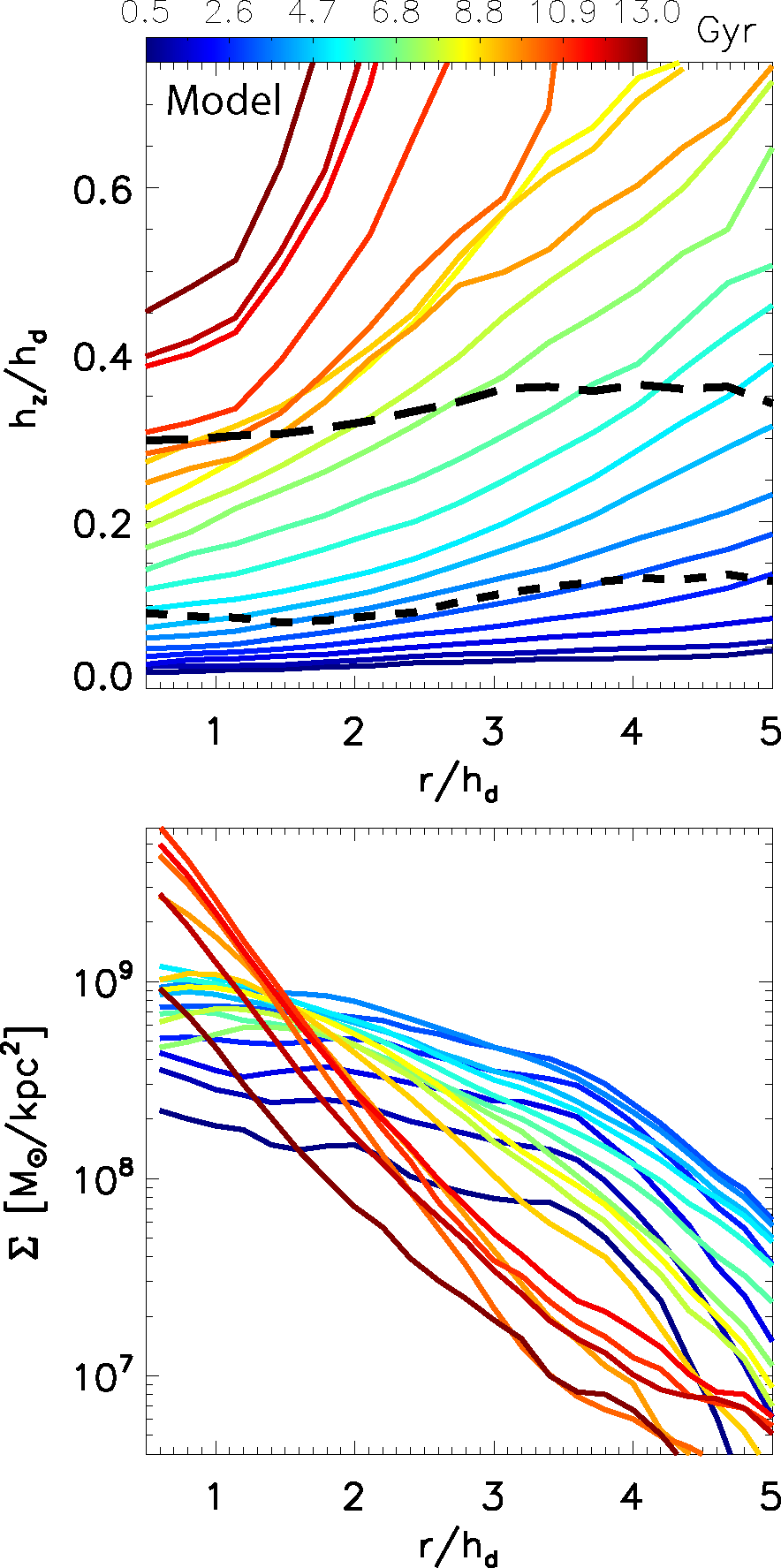}
\caption{
{\bf Top:} Variation of disk scale-height, $h_z$, with galactic radius for a cosmological simulation (see text).
Color lines show mono-age populations, as indicated. Bins of width one~Gyr are used and exponential models at different radii are fited out to $|z|=h_d$, where $h_d$ is the disk scale-length. Overlaid also are the thin (short-dash curve) and thick (long-dash curve) disks obtained by fitting a sum of two exponentials to stars of all ages. No significant flaring is found for the thin and thick disks. 
{\bf Bottom:} Disk surface density for mono-age populations showing that older samples are confined to the inner disk as a result of the inside-out disk formation. Flaring is lost in the total population since younger disks dominate in density at progressively larger radii. 
}
\label{fig:flaring}
\end{figure}

\section{Examples of Simpson's paradox}

\subsection{Variation of disk thickness with Galactocentric radius}
\label{sec:flaring}

As a first illustration of YSP, we consider a cosmological zoom-in hydrodynamical simulation, using initial conditions from one of the Aquarius Project haloes \citep{springel08, scannapieco09} and a Tree-PM SPH code. The spatial and mass resolutions are 300~pc and $4.4\times10^5$~M$_{\odot}$, respectively. The stellar mass is $5.5\times10^{10} M_{\odot}$ and the disk scale-length is $h_d=4$~kpc. More details about this simulation can be found in \cite{aumer13b} (model Aq-D-5).

In the top panel of Fig.~\ref{fig:flaring} color curves show the variation of disk scale-height, $h_z$, with galactic radius, $r$, for mono-age populations of width 1~Gyr and median values as seen in the color bar. Single exponential models were fitted to get $h_z$ in radial bins of width $0.5h_d$ and distance from the disk midplane $h_d$ (see details in \citealt{minchev15}). Strong disk flaring (an increase of scale-height with radius) can be seen for the oldest mono-age populations, decreasing for younger subsamples. When the total stellar population is considered, the vertical disk density at all radii requires a sum of two exponentials for a proper fit \citep{gilmore83}. Intriguingly, the resulting thin (short-dash curve) and thick (long-dash curve) disks do not show much flaring, even though they are composed of the strongly flared mono-age subsamples. This is an example of geometrical (also referred to as structural or morphological) definition of galactic thin and thick disks, as opposed to separation by age or by chemistry in the [$\alpha$/Fe]-[Fe/H] plane for which "low-" and "high-[$\alpha$/Fe] sequences", respectively, are more appropriate names (see discussion in \citealt{martig16b}).

This figure highlights a simple example of a "weak" case of YSP, where the strong variation in $h_z$ with radius for mono-age subsamples is lost when the total population is considered. As discussed by \cite{minchev15}, this results because the flaring of younger samples typically dominate at larger radii due to the inside-out disk formation. These "nested flares" constitute a geometric thick disk which does not flare, or in which the flaring is strongly reduced, consistent with the lack of disk flaring found in observations of external galaxies \citep{vanderkruit82, degrijs98, comeron11}. It can be seen in the bottom panel of Fig.~\ref{fig:flaring} that, indeed, the density of older mono-age groups drops much faster with radius and it flattens for younger populations.

\begin{figure*}
\centering
\includegraphics[width=0.8\linewidth, angle=0]{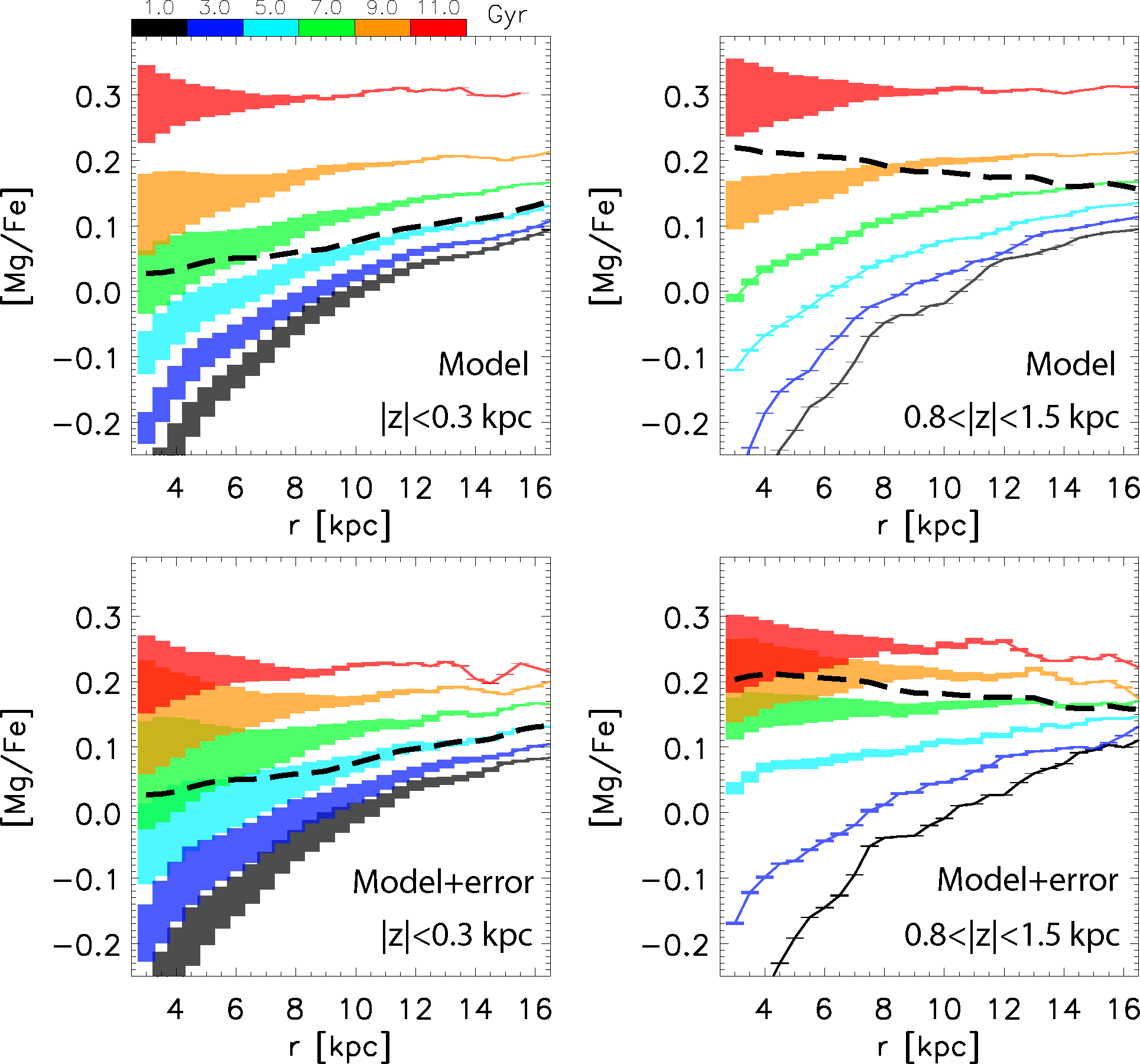}
\caption{
$\bf Top:$ Variation of azimuthally averaged [Mg/Fe] gradient with distance from the disk midplane for the MCM13 model. Dashed black curves show the [Mg/Fe] variation with Galactic radius for stellar samples at two different distances from the disk midplane, $|z|$, as marked in each panel. Different colors correspond to different age groups, as indicated in the color bar. The height of the colored rectangles reflects the stellar surfaces density of each radial bin. The positive gradient seen in the total population close to the plane (left panel) is reversed at high $|z|$ (right panel). The gradient of each individual age-bin, however, is positive or flat (for the oldest age group). 
{\bf Bottom:} Same as above, but with convolved uncertainties as expected from precise observations: $\rm\delta[Mg/Fe]=0.05$~dex and $\rm\delta age=20\%$. 
This figure illustrates both a weak (left column) and a strong case (right column) of Simpson's paradox.
}
\label{fig:mgfe}
\end{figure*}

This model of thick disk formation predicts a negative radial age gradient in geometrically-defined thick disks, which was found to be indeed the case for the Milky Way using APOGEE data \citep{martig16b}. Flaring in Milky Way stellar populations of limited age range has been found by \cite{kalberla14} (Red Clump), \cite{feast14} (Cepheids), \cite{carraro15} (young clusters) and in LAMOST \citep{wan17,wang18, xiang18} and APOGEE \citep{mackereth17} mono-age populations. A proper Milky Way disk mass model is needed to asses the interplay among the flares of different mono-age populations on the overall scale-height variation with radius (that of the total disk mass).

Flaring in mono-age populations from a theoretical point of view appears to be unavoidable, shown to result from merger perturbations \citep{kazantzidis08, villalobos08, martig14a}, misaligned gas infall \citep{scannapieco09, roskar10}, and reorientation of the disk rotation axis \citep{aumer13a}. A thick disk model incorporating disk flaring is attractive because flaring becomes a necessity rather than a nuisance (as often seen in numerical works in the past).

As will be shown next, inside-out disk formation combined with disk flaring can also explain the inversion of chemical abundance gradients with distance from the disk midplane.

$\bullet$ {\bf Possible erroneous interpretation in the absence of age measurements:} The Milky Way disk presents no flaring and thus dynamical mechanisms known to flare disks were not at work during its evolution. This is not true if you slice by age - a better interpretation is that mono-age disks always flare but flaring is lost in the total (or wide age range) sample due to the inside-out forming disk. This is a weak YSP case.

\begin{figure*}
\centering
\includegraphics[width=0.8\linewidth, angle=0]{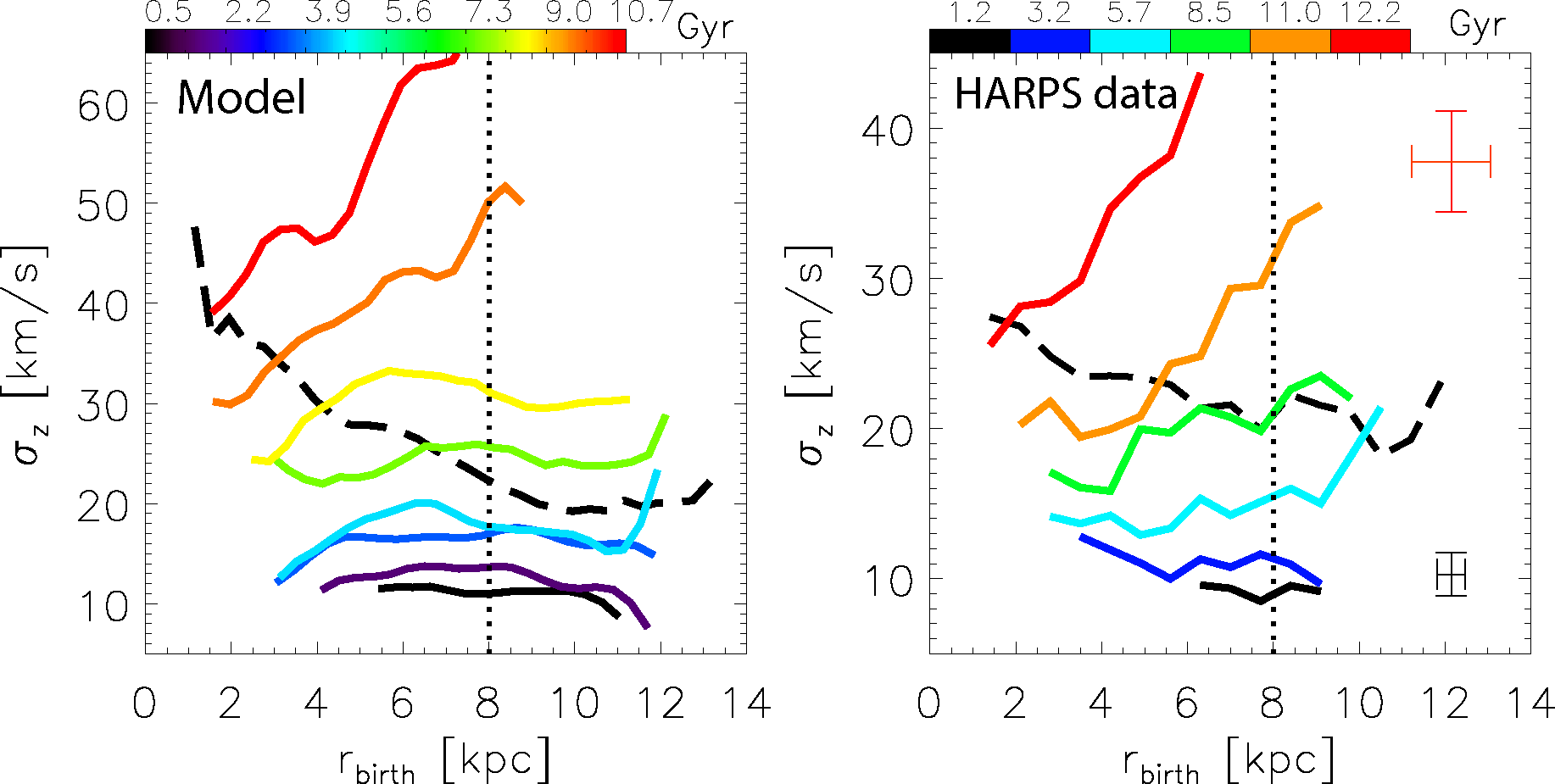}
\caption{
{\bf Left:} Predicted vertical velocity dispersion, $\sigma_z$, as a function of mean birth radius, $r_{birth}$, for the simulated solar neighborhood stellar sample of the MCM13 model ($d<0.5$~kpc). The black-dashed line shows the total population, suggesting that the hottest stars were born in the inner disk. Curves of different colors correspond to different age groups, with medians shown in the color bar. For the oldest samples stars arrive at the solar vicinity much cooler than stars born in-situ, due to the stronger effect of mergers on the outer disk and the decreasing probability of migration with increasing velocity dispersion. The large positive gradient found for old populations (red colors) turns negative for the youngest samples (blue and black), indicating a quiescent regime, where stars arriving from the inner disk heat slightly the local velocity distribution. {\bf Right:} Same as top but resulting from HARPS-GTO dwarf stars (ages from \citealt{anders18}, birth radii from \citealt{minchev18}). This plot, resulting from observations, is remarkably similar to the model shown in the left panel.
}
\label{fig:rbsigz}
\end{figure*}

\subsection{Inversion in radial abundance gradients}
\label{sec:grad}

As discussed in \S\ref{sec:intro}, inversion in the [$\alpha$/Fe] (e.g., [Mg/Fe]) gradient with distance from the disk midplane has been found in a number of Galactic surveys. We investigate the origin of this in the MCM13 model.

The top left panel of Fig.~\ref{fig:mgfe} shows the radial [Mg/Fe] gradient for the MCM13 model (similar to Fig.~10 by \citealt{mcm14}). The black-dashed curve results from the total stellar population with a maximum distances from the disk midplane of $|z|=0.3$~kpc. Different colors correspond to different mono-age populations, as indicated. Although the younger mono-age groups have well-defined positive slopes, that of the total population is strongly flattened inside $r\lesssim8$~kpc. This can be understood if we consider the relative stellar surface density as a function of radius for each mono-age population, which is indicated by the height of the colored rectangles (conveying the same information as the disk surface density variations with radius shown in the bottom panel of Fig.~\ref{fig:flaring}). The old stars -- red and orange bins -- dominate in the inner disk, causing the mean [Mg/Fe] at $r\lesssim8$~kpc to go up. On the other hand, the younger populations dominate in the outer disk, as evident from the total mean closely following the $\rm age=5$~Gyr line.

The top right panel of Fig.~\ref{fig:mgfe} is similar to the left one, but for a sample at $0.8<|z|<1.5$~kpc. For this significantly higher distance from the disk midplane, the density of old stars is found to strongly dominates at small radii, causing an increase in mean [Mg/Fe] by about $0.2$~dex compared to the sample confined to the disk plane and, thus, a negative gradient.

The bottom row of Fig.~\ref{fig:mgfe} presents the effect of uncertainties, as expected from high-resolution spectroscopy (e.g., APOGEE - \citealt{majewski17}, WEAVE - \citealt{dalton12}, 4MOST - \citealt{dejong12}) $\rm\delta[Mg/Fe]=0.05$~dex and asteroseismic ages (e.g., CoRoT, K2 - \citealt{howell14}, TESS - \citealt{ricker15} and PLATO - \citealt{rauer14}) $\rm\delta age=20\%$, drawn from a Gaussian distribution. Interestingly, the gradient of the total sample is only minimally affected, while a decrease of $\sim0.05$~dex is found in the oldest mono-age group.

The above illustrates a "strong" YSP case, resulting from the effect of inside-out disk formation combined with disk flaring (see \S\ref{sec:flaring}), giving rise to a different mixture of ages as a function of Galactic radius at different $|z|$ -- older, [$\alpha$/Fe]-rich stars are confined to the inner disk and large $|z|$, while younger, [Mg/Fe]-poor stars dominate above the disk plane in the outer disk. Similar explanation holds for the inversion of [Fe/H] gradient from negative close to the disk plane, to positive above it.

$\bullet$ {\bf Possible erroneous interpretation in the absence of age measurements:} 
(1) The sign shift in the [$\alpha$/Fe] and [Fe/H] radial gradients found in observations at larger distance form the disk midplane resulted from an outside-in (thick) disk formation or a dominating accreted population at large $|z|$. A better interpretation, according to our model, may be that [Mg/Fe] ([Fe/H]) depends negatively (positively) on $r$ only when one marginalizes over age. This is a strong YSP case. (2) The flattened gradient close to the disk plane is in disagreement with classical chemical evolution models. Steeper gradients are expected in mono-age populations. This is a weak YSP case.

\subsection{Birth radius vs vertical velocity dispersion}
\label{sec:rb}

We now demonstrate that disk flaring and heating as a function of radius can be seen in the local velocity distribution thanks to the effect of stellar radial migration.

In the left panel of Fig.~\ref{fig:rbsigz} we show the variation of the vertical velocity dispersion, $\sigma_z$, with birth radius, $r_{birth}$, for stars found in the simulated solar neighborhood of the MCM13 model. $\sigma_z$ is computed as the standard deviation of the vertical velocity in each radial bin. The black-dashed curve shows the total population, giving the impression that the hottest stars in the solar vicinity today have arrived from the inner disk. Looking at this relation it may be tempting to conclude that it is radial migration of hot inner disk stars that give rise to a thick disk. However, dissecting by age (color-coded curves) an opposite relation is revealed for the oldest stellar populations showing that stars born at 2-3 kpc have half the velocity dispersion of coeval populations born at the solar radius. This predicts that the kinematically hottest stars near the Sun today were born locally or in the outer disk - a natural result of the flaring in mono-age disks (see \S\ref{sec:flaring}), which is unavoidable in simulations for different reasons, the most likely possibly being perturbations from orbiting satellites.

The strong positive slope seen in old stars flattens toward redshift zero (present time), becoming slightly negative for age $\lesssim4$~Gyr, in agreement with the expected effect of migration in quiescent disks \citep{minchev12b,roskar13}. Note that the flaring induced by outward migration of stars with larger vertical actions and inward migration of outer-disk-born-stars with small vertical actions ($\sim50\%$ increase in $h_z$ in $\sim4$ disk scale-lengths, see \citealt{minchev12b}) is trumped by the impact of mergers and the effect of a cosmological context in general, in which case flaring results from locally born stars, i.e., non-migrators.

\begin{figure*}
\centering
\includegraphics[width=0.8\linewidth, angle=0]{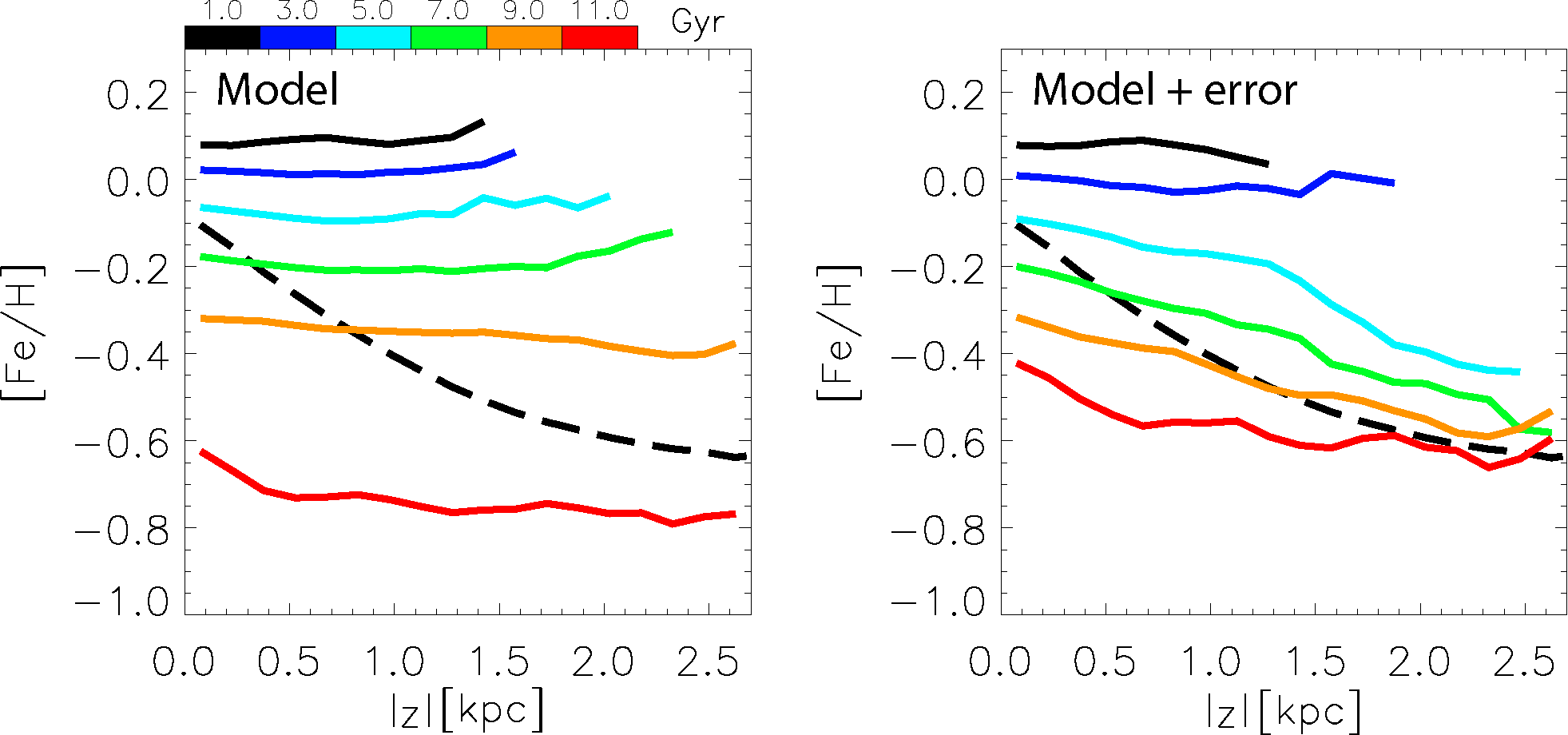}
\caption{
{\bf Left:} Mean [Fe/H] as a function of distance from the disk midplane, $|z|$, for stars in the simulated solar vicinity of the MCM13 model. The strong negative trend seen in the total population (black-dashed curve) is consistent with SEGUE G-dwarf data (see Fig.~14 in MCM13). In contrast, mono-age populations show much weaker variations. 
{\bf Right:} Same as left, but including uncertainties of $\rm\delta age=20\%$ and $\rm\delta[Fe/H]=0.1$~dex, considering a Gaussian probability distribution. While the mean of the total sample is not much affected, a notable negative gradient has appeared for mono-age population older than $\sim4$~Gyr (cyan through red curves).
}
\label{fig:fez}
\end{figure*}

The $r_{birth}-\sigma_z$ relation was first presented in order to understand why the velocity dispersion increase with [Mg/Fe] in RAVE Giants and SEGUE G-dwarfs showed an abrupt drop at the highest [Mg/Fe] end \citep{minchev14}. It was found that the observations could be explained by kinematically cool, old (thus low metallicity and high [Mg/Fe]) stars arriving to the solar neighborhood from low Galactic radii ($\sim2-4$ kpc). Indeed, larger changes in guiding radii (or angular momenta) are expected for the coldest subpopulations in a given age group (e.g., \citealt{vera-ciro14, daniel18}).  

\cite{minchev18} recently presented a largely model-independent method for estimating stellar birth radii based on age and [Fe/H] measurements, using AMBRE:HARPS \citep{delaverny13,depascale14, hayden17} and HARPS-GTO data \citep{adibekyan12, delgadomena17} combined with stellar ages computed with the StarHorse code \citep{santiago16, queiroz18}, as described in \cite{anders18}. This opened the possibility to test the $r_{birth}-\sigma_z$ relation in observations. 

In the right panel of Fig.~\ref{fig:rbsigz} we show $\sigma_z$ vs $r_{birth}$ using HARPS-GTO data for six mono-age populations with median values indicated in the color bar and a bin width $\rm \Delta age=2$~Gyr. Also shown are typical error bars, corresponding to two standard deviations of 1000 realizations in a bootstrapping calculation. As in \cite{minchev18}, we imposed the following quality criteria on ages and abundances: $\rm\delta[Mg/Fe]<0.07$~dex, $\rm\delta age/age<0.25$ or $\rm\delta age<1$~Gyr, resulting in a sample of 603 stars.

The trends in the data seen in the right panel of Fig.~\ref{fig:rbsigz} are remarkably similar to the MCM13 model expectation shown in the left panel. The result that stars born at larger radii are kinematically hotter than those born at smaller radii is remarkable as it provides, for the first time, observational evidence that outward migration cools the local disk, rather than heating it, as has been proposed in the past (e.g., \citealt{schonrich09b, roskar13}). This finding supports work showing that migration does not contribute to thick disk formation (e.g., \citealt{minchev12b, mcm14, vera-ciro14, grand16}, see also discussion in \S\ref{sec:intro}), except in redistributing stars heated by external mechanisms \citep{quinn93, villalobos08} or turbulent gas clouds at high redshift \citep{bournaud09,forbes12}.

$\bullet$ {\bf Possible erroneous interpretation in the absence of age measurements:} The trend in the total population of the $r_{birth}-\sigma_z$ relation suggests that migration from the inner disk created the (local) Milky Way thick disk. This is not true if we slice by age, finding exactly the opposite trend for old and intermediate ages in both model and data. A better interpretation is that hotter stars arrived from outer radii due to disk flaring. The outward migrators kinematically cool the local disk. This is a strong YSP case.

\subsection{The vertical metallicity gradient}
\label{sec:fez}

The left panel of Fig.~\ref{fig:fez} presents the mean [Fe/H] as a function of distance from the disk midplane, $|z|$, for stars in the simulated solar vicinity of the MCM13 model. A negative trend in the total population (black-dashed curve) is seen, consistent with SEGUE G-dwarf \citep{schlesinger12, rix13} and RAVE data \citep{boeche13b}. In contrast, mono-age populations (color-coded curves) show much weaker variations - some decrease for old and an increase for the two youngest age bins. 

In the right panel of Fig.~\ref{fig:fez} we show the effect of observational uncertainties, considering $\rm\delta age=20\%$ and $\rm\delta[Fe/H]=0.1$~dex and a Gaussian probability distributions (as in Fig.~\ref{fig:mgfe}). Similarly to the radial [Mg/Fe] gradients (\S\ref{sec:grad}), the observational error does not seem to affect much the total sample, however, notable negative trends appear for mono-age populations older than $\sim4$~Gyr (cyan through red curves). A remarkable drop of $\approx0.3$ and $\approx0.4$ dex at high $|z|$ is seen for the green and cyan curves, respectively. Recently, \cite{ciuca18} found similar negative trends in the older populations in RAVE DR5 \citep{kunder17} data, which, given the results of this section, may be caused by observational uncertainties.

The explanation for the negative vertical metallicity gradient in the total population is again related to the interplay among stellar ages, chemical enrichment, and stellar density, similarly to the radial [Mg/Fe] gradient inversion with $|z|$ (\S\ref{sec:grad}). Younger metal-rich mono-age groups dominate closer to the disk midplane, while older metal-poor ones are more abundant at larger $|z|$. The fraction of younger to older stars varies with Galactic radius. This both causes a variation in the vertical metallicity gradient (see, e.g., \citealt{hayden14} using APOGEE data) and an inversion of [Fe/H] and [Mg/Fe] gradients with increasing $|z|$ discussed in \S\ref{sec:grad}. Depending on the observational biases of different dataset, different slopes in [Fe/H] (or [M/H]) with $|z|$ would be measured in the total population as summarized by \cite{schlesinger14}, which may result in different interpretations.

\begin{figure*}
\centering
\includegraphics[width=0.9\linewidth, angle=0]{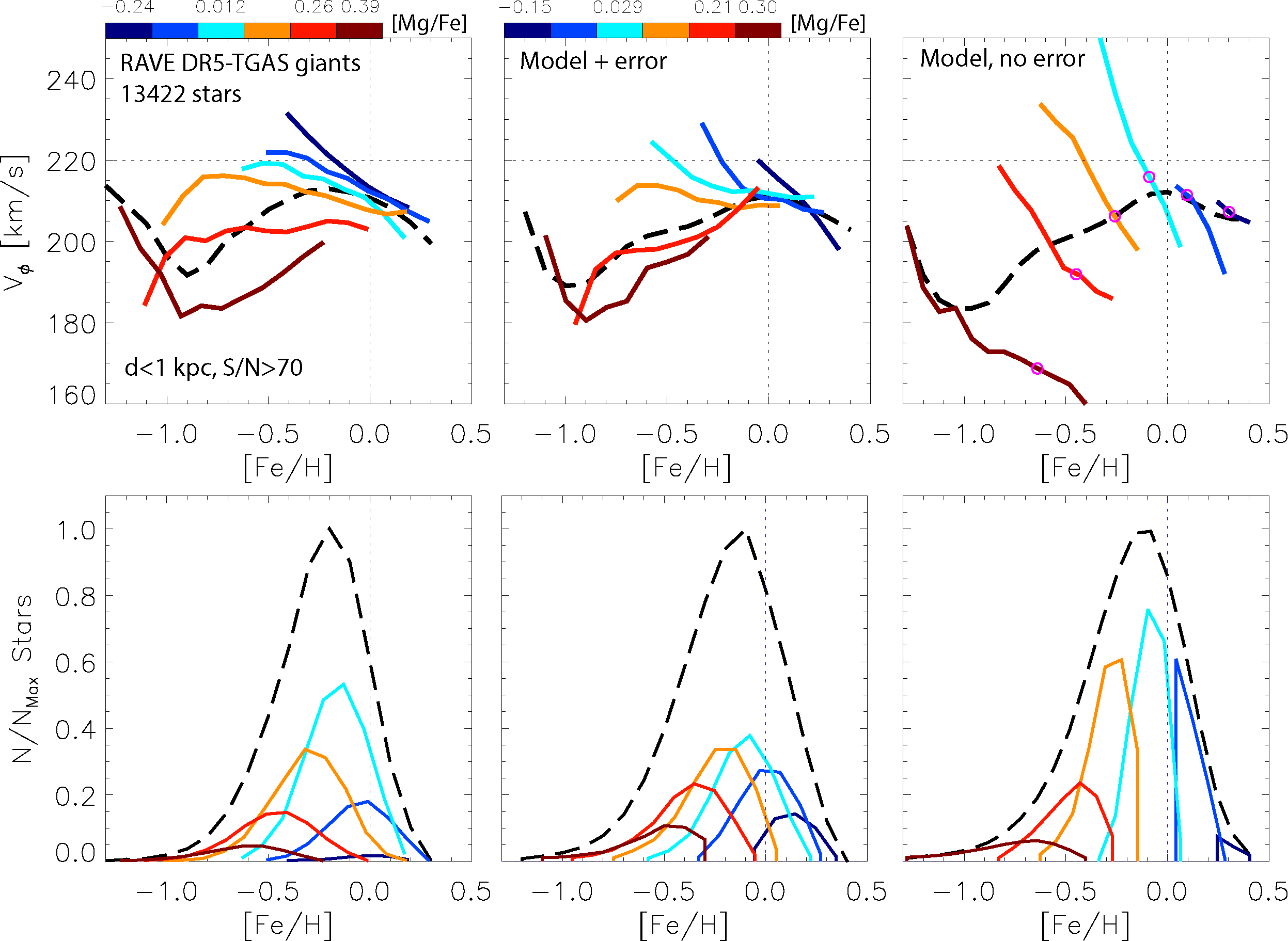}
\caption{
Comparison between the metallicity-rotational velocity relation (MVR) in RAVE-TGAS giants and the MCM13 model. 
{\bf Left column:} RAVE-TGAS data with S/N$>70$ and $d<1$ kpc. The black-dashed curve in the top panel shows the rotational velocity as a function of metallicity for the total population, while color-coded solid curves correspond to different median [Mg/Fe] bins, as indicated in the color bar. The bottom panel shows the distribution of the total (black-dashed) and [Mg/Fe] sub-populations. Stars with $V_\phi<100$~km/s have been discarded in both data and model.
{\bf Middle column:} Same as left, but for the MCM13 model simulated neighborhood, including velocity and abundance uncertainties (see text). For both the data and model an S-shape in the total population MVR is seen, as well as a remarkably good match among different sub-populations.
{\bf Right column:} Same as middle, but {\it excluding} uncertainties. It can be seen that for each [Mg/Fe] sub-population the slope is negative. The MVR can become positive at [Fe/H]$<0$ if (1) a sample spans a large range of [Mg/Fe] or (2) uncertainties are not sufficiently small.
}
\label{fig:mvr}
\end{figure*}

\cite{ciuca18} reasoned that the mild negative gradient seen in the oldest population suggested mono-age disks must have been born flared, which after radial mixing would bring hotter, metal-poor stars from the outer disk to the solar neighborhood, for a given mono-age population. The effect will be the same, however, even if mono-age disks were heated subsequently by external perturbations, i.e., the same scenario producing the nested flares in Fig.~\ref{fig:flaring}. The slightly positive gradient in the youngest populations then signifies the opposite effect - migrators from the inner disk weakly heat the local disk, while those arriving from the outer disk cool it, mostly cancelling the overall contribution (as seen in Fig.~5 by \citealt{minchev12b} and in the negative slope of the youngest stars in Fig.~\ref{fig:rbsigz}). The latter effect results only in disk evolution lacking external perturbations. 

This section presents a weak YSP case.

\subsection{The metallicity-velocity relation (MVR)}
\label{sec:mvr}

A correlation between the observed total metallicity, [M/H], and rotational velocity in the Galactic plane, $V_\phi$, of stars in the solar neighborhood was already reported by \cite{carney90}. \cite{haywood08} suggested that this metallicity-velocity relation (MVR), can serve as a test for the amount of stars migrated into the solar neighborhood. He showed that metal-poor solar neighborhood stars at low [$\alpha$/Fe] possess large angular momenta. This meant that they were visitors from the outer disk, currently near the Sun on their pericenters. \cite{schonrich09a} argued that the presence of a negative correlation in the MVR for stars in the chemically defined thin disk (or more correctly, the low-[$\alpha$/Fe] sequence) is an indication of migration. Using a compilation of data for stars in the solar neighborhood, \cite{navarro11} found a flat MVR for low-[$\alpha$/Fe] stars, concluding that migration in the solar neighborhood was unimportant. \cite{lee11} found a negative relation for low-[$\alpha$/Fe] stars, in contrast to \cite{navarro11}. A positive slope in the MVR for high-[$\alpha$/Fe] stars was described by \cite{spagna10}, \cite{lee11}, \cite{kordopatis11}, and \cite{adibekyan13}. More recently, \cite{allende-prieto16} used the APOGEE-TGAS sample which utilizes the Gaia DR1 data \citep{gaia16}, confirmed the positive and negative slopes of the high- and low-[$\alpha$/Fe] stellar populations in the local disk, respectively. 

It will be shown below that the positive slope in the MVR of high-[$\alpha$/Fe] stars spanning a wide range in age is just another strong YSP case resulting from the combination of MVRs of mono-age (or mono-[Mg/Fe]\footnote{[$\alpha$/Fe] is a good proxy for age for samples confined to a small radial bin \citep{haywood13, bergemann14}, but not for data sets spanning large Galactic radii \citep{minchev17}.}) populations with well-defined negative slopes.\footnote{These results were first presented at the RAVE collaboration meeting in Ljubliana, Slovenia in 2014, using RAVE-DR4 \citep{kordopatis13b}.}

We use a subsample from RAVE DR5 \citep{kunder17} with high-quality chemical and kinematical information. We selected a sample of giant stars close to a random magnitude-limited sample as done in \cite{minchev14}. We excluded giants with $\rm log(g) < 0.5$ to avoid  possible effects due to the boundaries of the learning grid used for the automated parameterization, and considered stars in the temperature range $\rm 4000 < T_{eff} < 5500$~K (thus avoiding horizontal branch stars; see \citealt{boeche13a}). Our final sample consists of stars with signal to noise S/N$>70$ and a distance from the Sun $d<1$ kpc.

The top-left panel of Fig.~\ref{fig:mvr} presents the MVR for our RAVE-TGAS sample. The total population (black-dashed curve) shows a distinct S-shape rotated by $\sim90^\circ$. A positive trend is found in the range ${\rm-1<[Fe/H]<0}$, while a negative one can be seen at ${\rm[Fe/H]>0}$ and ${\rm[Fe/H]<-1}$~dex. When we consider narrow bins in [Mg/Fe] (color-coded curves), however, the MVRs of low-[Mg/Fe] groups show strong negative slopes, flattening and inverting as [Mg/Fe] increases. A V-shape is seen in the highest [Mg/Fe] bin. 

The top-middle panel of Fig.~\ref{fig:mvr} is similar to the left one, but showing the MCM13 model, including uncertainties similar to what we expect for the data: $\rm\delta[Mg/Fe]=0.12$~dex, $\rm\delta[Fe/H]=0.1$~dex, and $\rm\delta V_\phi=2$~km/s. A remarkable match is seen to the RAVE-TGAS data for both the total population and the subsamples, notably, the gradual inversion from negative to positive with increasing [Mg/Fe] and the V-shape of the highest [Mg/Fe] bin.

Finally, in the top-right panel of Fig.~\ref{fig:mvr} we present the model {\it excluding} the uncertainties, which shows a drastic change in shape for individual sub-populations. The MVR trend is always negative, slightly decreasing in steepness for the highest [Mg/Fe] bin. The slope measures $\rm dV_\phi/d[Fe/H]=-127$~km/s/dex for stars with median [Mg/Fe] of 0.13~dex (cyan curve), flattening to $\rm dV_\phi/d[Fe/H]=-50$~km/s/dex for the highest [Mg/Fe] bin with median value of 0.34~dex (maroon curve).

\subsubsection{Effect of observational uncertainties}

To understand the effect of observational uncertainties seen in the model, in the bottom row of Fig.~\ref{fig:mvr} we show the [Fe/H] distributions of the total samples and the corresponding mono-[Mg/Fe] populations. In both data and model we see similar shifts in the peaks to lower [Fe/H] with increasing [Mg/Fe], as expected. 

The small pink circles in the top-right panel of Fig.~\ref{fig:mvr} indicate the approximate [Fe/H] values at which the density peaks for each mono-[Mg/Fe] population. We see that those act as pivot points around which the MVR flattens when neighboring bins are allowed to communicate, i.e., when uncertainties are convolved with the model (middle column). The only exception is the highest-[Mg/Fe] bin, because it is significantly lower in density than the intermediate bins, and it lacks a clear peak.

In a range of [Fe/H] over which the densities of two adjacent mono-[Mg/Fe] populations are comparable, both MVRs are deflected similarly. For example, the two highest-[Mg/Fe] bins in the range $-1.0<$[Fe/H]$<-0.5$~dex show similar densities, consequently, the maroon curve shifts upwards and red curve downwards as error is implemented. More generally, the MVRs of density tails are affected by the neighboring bin's peaks in the corresponding direction: since the slope is always negative, metal-poor tails are shifted toward larger $V_\phi$ and vice versa. It should be noted that relatively similar densities in the observations are not to be taken at face value due to the selection function of a particular survey. For example, the RAVE survey is known to favor low-[$\alpha$/Fe] stars \citep{boeche13b}, which is evident in the distributions of mono-[Mg/Fe] populations in the bottom left panel of Fig.~\ref{fig:mvr}.

The negative MVR slope at [Fe/H]$<-0.9$~dex is minimally affected, as these are the stars with the highest [Mg/Fe] values and the uncertainties are not large enough to allow other [Mg/Fe] bins to reach that end of the metallicity distribution. This distinct shape is present in both data and model, as well as seen in other works, although not acknowledged. For example, such an inversion can be found (1) in the bottom panel of Fig.~7 by \cite{lee11}, if one focussed on the mean values, rather than the line fit, and (2) in the top panel of Fig.~4 by \cite{adibekyan13} at $\rm [Fe/H]\approx -1.1$~dex.

Similarly to the effect of errors, using wider [Mg/Fe] ranges would result in overlapping between the neighboring bins (all of which show clearly negative MVRs), approaching the black-dashed line in the right panel of Fig.~\ref{fig:mvr}. If we split into the high- and low-[Mg/Fe] sequences, as often done in the literature, we would end up with a positive and a negative MVR, respectively.

In summary, the genuinely negative MVR can become positive if (1) a sample spans a large range of [Mg/Fe] or (2) uncertainties are not sufficiently small.

\begin{figure}
\centering
\includegraphics[width=0.8\linewidth]{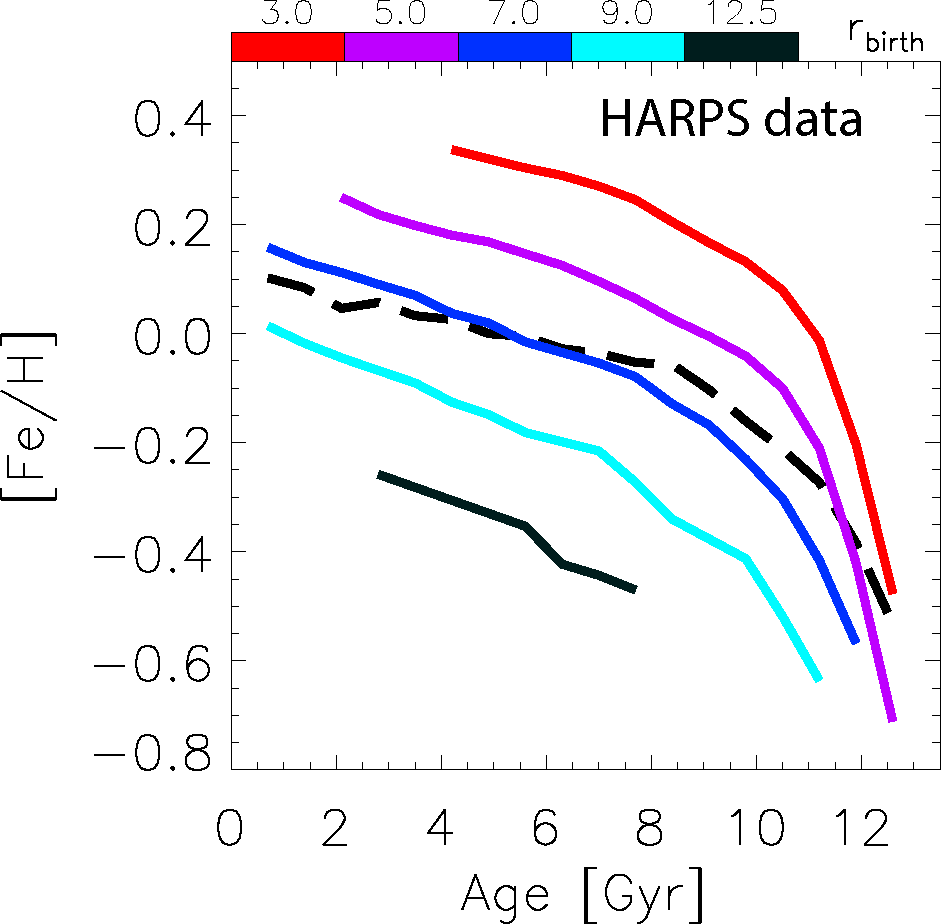}
\caption{  
The local age-metallicity relation (AMR) using our local HARPS-GTO data. The black-dashed curve shows the mean resulting from the total population. The color-coded lines represent five mono-$r_{birth}$ populations, with median values indicated in the color bar and a bin width $\Delta r_{birth}=2$~kpc. Well-defined AMRs exist for stars with common birth radii. The AMR is shifted downward for outer radii, which is a result of the negative ISM radial metallicity gradient. Note that there is no galactic model involved here.
}
\label{fig:amr}
\end{figure}

\subsubsection{What causes the invariantly negative MVR?}

The negative trends for all mono-[Mg/Fe] populations (right panel of Fig.~\ref{fig:mvr}) are easy to understand. While a star at the solar radius, $r_0$, moving on a perfectly circular orbit rotates with the value of the circular velocity (here $V_0=220$ km/s), stars with guiding radii inside/outside $r_0$ will rotate slower/faster than that. Due to the exponential stellar surface density decrease with radius, however, more low-angular momentum stars are present in the solar neighborhood (this is the asymmetric drift effect, see \citealt{bt08}). Consequently, the mean tangential velocity, $V_\phi$, of the total population is always below $V_0$. Since our sample is selected from a narrow radial range, the angular momentum is given by $L\approx V_\phi r_0$ and in terms of the stellar guiding radius, $r_g$, as $L\approx V_0 r_g$. Therefore, $V_\phi\approx (V_0/r_0) r_g$, i.e., $V_\phi$ can be thought of as a proxy for $r_g$. We will not find the youngest stars with guiding radii far away from $r_0$, as those will be on orbits too close to circular to be able to reach our sample. The total population shows a peak at solar [Fe/H] because these are the youngest stars near the solar radius, thus having the most circular orbits. 

Splitting the model sample in narrow bins of [Mg/Fe] in the top-right panel of Fig.~\ref{fig:mvr} showed that the MVR is always negative. This simply reflects the radial negative metallicity gradient of a mono-age, or in this case, a mono-[Mg/Fe] population. Since $V_\phi\sim r_g$, for a given [Mg/Fe] bin, a negative metallicity gradient is reflected into a negative MVR gradient. 

In summary, the positive MVR gradient in the high-[$\alpha$/Fe] stars measured in a number of works \citep{carney90,rocha-pinto06,spagna10,lee11,adibekyan13,allende-prieto16} is simply the result of the asymmetric drift and the negative radial metallicity gradient of mono-age populations.

$\bullet$ {\bf Possible erroneous interpretation in the absence of age measurements:} The MVR slope of a mixed-age population tells us something about the Milky Way migration history. In fact the MVR slope is strongly dependent on the sample selection function and the age range of the sample. When dissected by age (or [$\alpha$/Fe]), one finds that all groups have negative slopes closely related to the radial metallicity gradients of mono-age populations. This is a strong YSP case.

\subsection{The age-metallicity relation (AMR)}
\label{sec:amr}

The age-metallicity relation (AMR) in the solar vicinity is believed to give us a measure of how fast the iron content of stars has increased with cosmic time. It has been found somewhat confusing that the AMR is quite flat, or even reversed, for stars with $\rm age\lesssim10$~Gyr (e.g., \citealt{edvardsson93, lin18,vickers18}). This has been taken sometimes as evidence that stars at the solar radius have not enriched in iron during that period, which is in disagreement with predictions from Milky Way chemical evolution models, or thought to result from the target selection of a given sample. While it is widely accepted now that radial migration is responsible for the scatter in [Fe/H] at the same age \citep{sellwood02,roskar08a, schonrich09a}, it has only recently been suggested that stars migrating from different radii present well-defined AMRs but conspire to flatten the relation resulting from the total population \citep{minchev18}.

In Fig.~\ref{fig:amr} we plot the AMR for the local HARPS-GTO data, which was used in the right panel of Fig.~\ref{fig:rbsigz}. The black-dashed curve shows the relation for the total sample. The color-coded lines represent five mono-$r_{birth}$ populations, as color-coded, of width 2~kpc and median values given in the color bar. Birth radii were estimated from the stellar age and [Fe/H], as described by \cite{minchev18}. This is not an unbiased result since the $r_{birth}$ estimate is based on age and [Fe/H]. Nevertheless, this figure reflects how the functional form of the ISM metallicity evolution with radius and cosmic time (also estimated by \citealt{minchev18}) determines the shape of the mono-$r_{birth}$ AMRs.

\begin{figure*}
\centering
\includegraphics[width=0.8\linewidth]{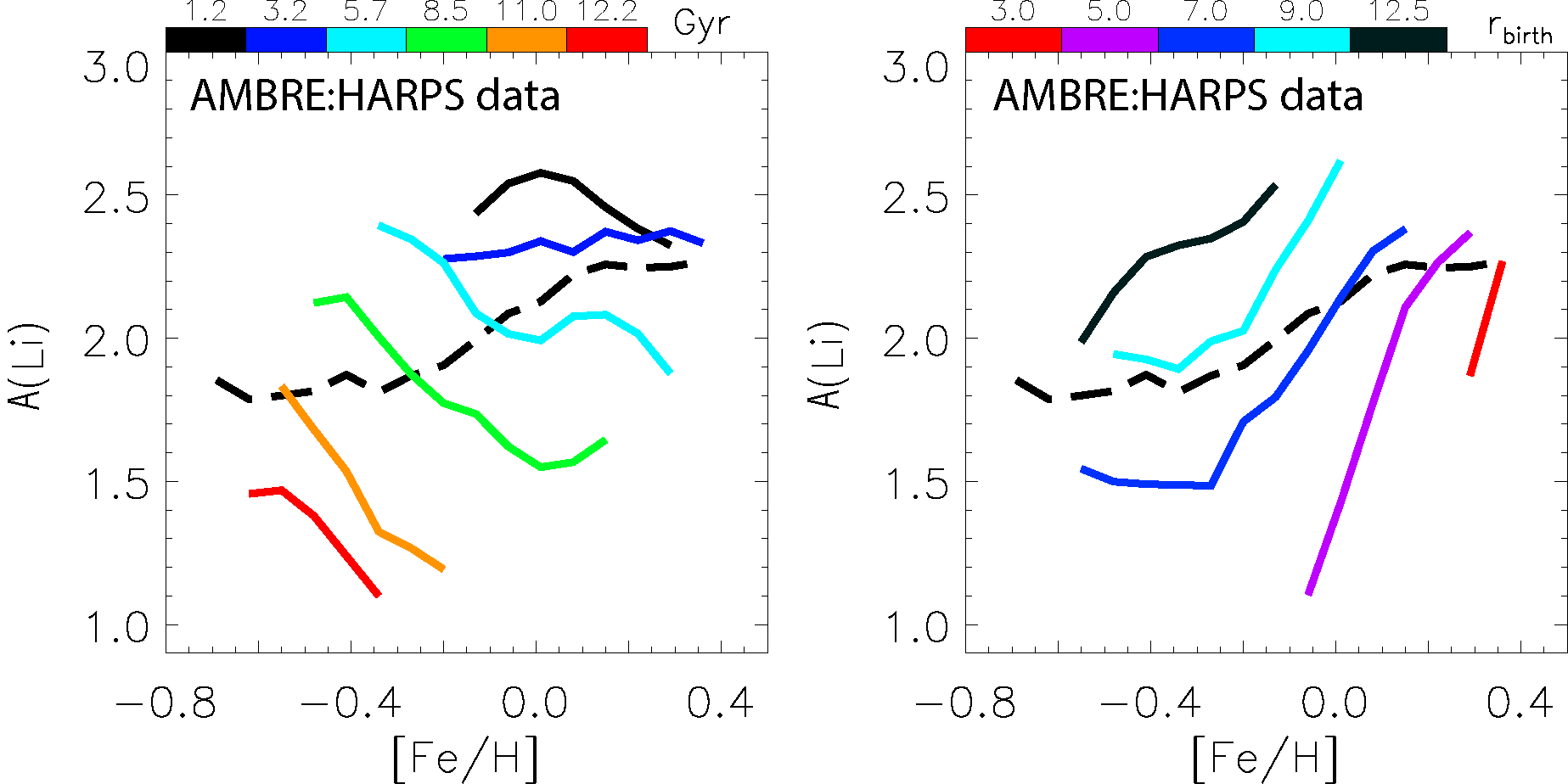}
\caption{  
{\bf Left:} Lithium abundance, A(Li), vs [Fe/H] for the local AMBRE:HARPS data \citep{guiglion16}. The black-dashed curve shows the mean resulting from the total population. The color-coded lines represent six mono-age populations, with median values indicated in the color bar and a bin width 2.5~Gyr. Mono-age populations show typically negative trends while the total populations has a positive slope.
{\bf Right:} Same as left, but color curves show mono-$r_{birth}$ populations. As in the case of the AMR, steeper slopes are found for stars coming from similar radii, compared to the total sample. This figure presents a strong (left panel) and a weak (right panel) YSP cases.
}
\label{fig:li}
\end{figure*}

In Fig.~\ref{fig:amr} we measure a gradient of 0.02 dex/Gyr for $\rm age<8.5$~Gyr in the total population (black-dashed curve). However, the slope in the AMR of stars born in the range $6<r_{birth}<8$~kpc (blue curve) is 0.04 dex/Gyr and is similar for the other $r_{birth}$ bins. This figure provides an explanation for the flatness of the local AMR, when radial migration is not taken into account. The slope of the total population will depend crucially on selection effects of a particular survey and the quality cuts applied. In this particular case we find quite a steep one compared to other data sets. For example, from Fig.~7 by \cite{lin18} we estimated $\sim0.013$~dex/Gyr for $\rm age<10$~Gyr, while \cite{vickers18} find a double peak at $\sim3$ and $\sim8$~Gyr. While the shape indeed depends on the particular target selection when the total sample is considered \cite{casagrande16}, such effects will be significantly reduced if we split in mono-$r_{birth}$ populations, relying on the assumption that stars with a particular value of [Fe/H] and age were born at roughly the same radius. Therefore, we do not expect the AMRs of mono-$r_{birth}$ populations to deviate much in other data sets from what we show here.

In summary, the chemical enrichment and the radial migration history are so closely entwined in the data that they can only be inferred together, as showed by \cite{minchev18} and \cite{frankel18}.

$\bullet$ {\bf Possible erroneous interpretation in the absence $r_{birth}$ measurements:} The metallicity at the solar radius has not increased in the last 8-10 Gyr of evolution. This is not true if mono-$r_{birth}$ populations are considered, showing well defined slopes in the AMR of stars born at a given radial bin. This is a weak YSP case.

\subsection{Lithium abundance variation with [Fe/H] }
\label{sec:li}

The left panel of Fig.~\ref{fig:li} shows the abundance of lithium, A(Li), vs [Fe/H] for our local HARPS data used in \S\ref{sec:rb} and \S\ref{sec:amr}. The difference here is that the lithium is estimated from the AMBRE:HARPS sample (as detailed by \citealt{guiglion16b,guiglion16}) and crossmatched with the ages derived from HARPS-GTO stellar parameters (\citealt{anders18}, using the StarHorse code), resulting in 326 dwarf stars. The black-dashed curve shows the mean resulting from the total population and the color-coded lines represent six mono-age populations, as in Fig.~\ref{fig:rbsigz}. While the relation from the total sample has a positive slope, mono-age populations show negative trends, except for the two youngest age bins. The variation of mean A(Li) with [Fe/H] for mono-age groups, to our knowledge, has not been shown before in observations, although it could be argued would result from the scatter plots shown in Figures~5 and 7 by \cite{ramirez12} and should correspond to bins of common stellar mass, shown in several works (e.g., \citealt{nissen12, ramirez12, delgadomena15, bensby18}). The flattening we find in A(Li)-[Fe/H] for younger groups of stars is very similar to the flattening seen with increasing stellar mass in Fig.~6 by \cite{bensby18}, as expected from the correspondence between stellar mass and age. This is, once again, a clear strong case of YSP, akin to the $r_{birth}-\sigma_z$ relation of \S\ref{sec:rb} and the MVR of \S\ref{sec:mvr}, which all show a dramatic reversal of the trend in the total local population.

Studies are usually concerned with the {\it upper envelope} (not shown here) in the A(Li)-[Fe/H] relation of dwarf stars, reasoning that this is closer to the original lithium abundance in the ISM, since depletion over the lifetime of stars takes place (e.g., \citealt{rebolo88, romano99, lambert04, guiglion16, cescutti19, guiglion19}). This {\it upper envelope}, which will be traced by extending the metal-poor tails of the mono-age populations in Fig.~\ref{fig:li}, is found to delineate stars from old to young when moving from low to high [Fe/H]. Further work is needed to understand if this is simply due to the inability of young stars to reach the solar neighborhood on apo- and peri-centers as their orbits are close to circular, and the insufficient time for migration.

Splitting the same data by birth radius (right panel of Fig.~\ref{fig:li}), we find that mono-$r_{birth}$ populations exhibit mostly positive trends as in the total sample, but with significantly steeper slopes (a weak YSP case, similar to the AMR relation, see Fig.~\ref{fig:amr}). This general shift from high to low [Fe/H] for inner to outer $r_{birth}$ bins is also seen in the chemical evolution models by \cite{prantzos17} and \cite{chiappini09a} (shown by \citealt{guiglion19}). In those models, however, the variation of A(Li) with [Fe/H] is shown for the gas, i.e., at the time of stellar birth. Since lithium is depleted throughout stars' lifetime, we need to consider the {\it upper envelope} for each $r_{birth}$ bin. The above models provide a good match to the trends seen in the right panel of Fig.~\ref{fig:li} assuming the scatter around the mean for all mono-$r_{birth}$ groups is similar. 

\cite{delgadomena15} described a lithium decrease at $\rm[Fe/H] > 0$ using HARPS data, a result subsequently confirmed by \cite{guiglion16} using a larger dataset. Recently \cite{guiglion19} reasoned that the observed decline in lithium abundance at super-solar metallicity can be explained as the effect radial migration, as stars in the metal-rich tail are expected to have been born roughly in the range $2<r<5$ kpc (see Fig.~8 by \citealt{minchev18}). The right panel of Fig.~\ref{fig:li} shows this is indeed the case, using our estimated birth radii for HARPS data.

It should be noted that our results are not dependent on any chemical evolution or dynamical modeling, except for a simple assumption on keeping the mono-$r_{birth}$ distributions physically meaningful (see for details \citealt{minchev18}). It is clear that, depending on the birth radius bin, the {\it upper envelope} will have different shape, notably, shifting to higher [Fe/H] and having a faster drop at the [Fe/H]-poor tail, as $r_{birth}$ decreases. This opens up the possibility to study the production/destruction of lithium as a function of birth radius, that will provide stronger constraints on chemical evolution models.

$\bullet$ {\bf Possible erroneous interpretation in the absence of age and $r_{birth}$ measurements:}
In the total sample the {\it upper envelope} in the A(Li)-[Fe/H] plane will be biased toward older stars at the higher and (especially) lower [Fe/H] tails (see Fig.~2 by \citealt{minchev18}), in which case the inferred ISM evolution of lithium will be incorrect. This problem can be resolved with the help of age and $r_{birth}$ estimates. Dissecting the A(Li)-[Fe/H] relation by age presents a strong YSP case and that by $r_{birth}$ - a weak one.

\section{Conclusions}

We showed in this work that Yule-Simpson's paradox (YSP) is omnipresent in the field of Galactic Archaeology. Chemo-kinematical relations believed to constrain the Milky Way past can look completely different, depending on whether a subset of stars with common age or a mixture of ages is considered (and similarly for birth radius). In other words, a statistical relationship between two variables does not necessarily represent a cause-and-effect relationship. 
We can summarize our results as follows:

$\bullet$ We described seven YSP cases, which we classified as "weak" if a relation was flattened or weakened, or "strong" if it was completely reversed. Weak cases were found in the variation of galactic disk thickness with Galactocentric radius (\S\ref{sec:flaring}), the vertical metallicity gradient (\S\ref{sec:fez}), the age-metallicity relation (AMR, \S\ref{sec:amr}), and the A(Li)-[Fe/H] relation when split by birth radius (\S\ref{sec:li}). Strong cases appeared in the inversion of radial abundance gradients with distance from the disk midplane (\S\ref{sec:grad}), the $r_{birth}-\sigma_z$ relation (\S\ref{sec:rb}), the metallicity-rotational velocity relation (MVR, \S\ref{sec:mvr}), and the A(Li)-[Fe/H] relation when split by age (\S\ref{sec:li}).

$\bullet$ YSP is found in both global (\S\ref{sec:flaring} and \S\ref{sec:grad}) and local (samples confined to the solar neighborhood, \S\S\ref{sec:rb}-\ref{sec:li}) chemo-kinematical trends.

$\bullet$ The shape of the MVR of stars in the solar vicinity has been suggested as a discriminant for the amount of radial migration suffered by the local disk, where a flat (inverse) relation has been argued to signify the unimportance (importance) of migration. Using RAVE data and the MCM13 chemodynamical model, we showed that the slope in the MVR for any narrow [Mg/Fe] subpopulation (used as a proxy for age) is always negative. An inversion from a negative to a positive slope of high-[$\alpha$/Fe] stars can be caused by the increasingly lower angular momentum of older, kinematically hotter stars at [Fe/H]$\lesssim-0.5$~dex with a slope measurement dependent on the selection function of a particular survey, as well as the observational uncertainties. The persistent MVR negative slope for any [$\alpha$/Fe] bin simply reflects the Milky Way negative metallicity gradient of mono-age populations due to the inside-out disk formation, combined with the effect of the asymmetric drift.

$\bullet$ Using estimates of stellar birth radii for AMBRE:HARPS data, we showed that the A(Li)-[Fe/H] relation in the solar neighborhood looks different for different mono-$r_{birth}$ populations. This can provide new constraints on chemical evolution models by requiring to match the {\it upper envelope} of the relation in different $r_{birth}$ bins, as well as their combination at a given final radius expected from the effect of radial migration.

Depending on the observational biases of different dataset, different trends can be measured in the total population of a given relation, thus resulting in various interpretations. Dissecting by age, however, decreases strongly such biases, as coeval stars will be affected by dynamical processes similarly, for a given radius. Moreover, one need not worry what age a particular stellar sample is biased toward, as long as the relation of interest does not involve the stellar density. For example, in the case of the vertical metallicity gradient, assuming stars with the same metallicity and age are born at the same radius (i.e., no significant intrinsic abundance scatter at the time of star formation, cf, \citealt{spitoni19}), the only effect we need to worry about is radial migration. 

We demonstrated that splitting a stellar sample by age or birth radius is of utmost importance for recovering the Galaxy evolution. Given that our birth radius estimate depends crucially on precise ages measurements, we conclude that ages are urgently needed to make progress in the field of Galactic Archaeology. We expect to get those form the K2 \citep{howell14}, TESS \citep{ricker15} and PLATO \citep{rauer14} asteroseismic missions in the near future, as well as isochrone age determination from spectroscopic surveys, such as GALAH \citep{desilva15}, Gaia-ESO \citep{gilmore12}, APOGEE, WEAVE, SDSS-V's Milky Way Mapper \citep{kollmeier17}, and 4MOST, combined with Gaia astrometry. 

Simpson's paradox undoubtedly exists in other Galactic Archaeology relations, as well as other areas of astrophysics. There may be other lurking variables in addition to age and birth radius we discussed here, which must be accounted for in statistical experiments, in order to avoid non-meaningful results. We hope this work brings awareness to different communities and helps advance our understanding of the Milky Way formation and evolution.


\bibliographystyle{mnras}
\bibliography{myreferences}

\begin{thebibliography}{}
\makeatletter
\relax
\def\mn@urlcharsother{\let\do\@makeother \do\$\do\&\do\#\do\^\do\_\do\%\do\~}
\def\mn@doi{\begingroup\mn@urlcharsother \@ifnextchar [ {\mn@doi@}
  {\mn@doi@[]}}
\def\mn@doi@[#1]#2{\def\@tempa{#1}\ifx\@tempa\@empty \href
  {http://dx.doi.org/#2} {doi:#2}\else \href {http://dx.doi.org/#2} {#1}\fi
  \endgroup}
\def\mn@eprint#1#2{\mn@eprint@#1:#2::\@nil}
\def\mn@eprint@arXiv#1{\href {http://arxiv.org/abs/#1} {{\tt arXiv:#1}}}
\def\mn@eprint@dblp#1{\href {http://dblp.uni-trier.de/rec/bibtex/#1.xml}
  {dblp:#1}}
\def\mn@eprint@#1:#2:#3:#4\@nil{\def\@tempa {#1}\def\@tempb {#2}\def\@tempc
  {#3}\ifx \@tempc \@empty \let \@tempc \@tempb \let \@tempb \@tempa \fi \ifx
  \@tempb \@empty \def\@tempb {arXiv}\fi \@ifundefined
  {mn@eprint@\@tempb}{\@tempb:\@tempc}{\expandafter \expandafter \csname
  mn@eprint@\@tempb\endcsname \expandafter{\@tempc}}}

\bibitem[\protect\citeauthoryear{{Adibekyan}, {Sousa}, {Santos}, {Delgado
  Mena}, {Gonz{\'a}lez Hern{\'a}ndez}, {Israelian}, {Mayor}  \&
  {Khachatryan}}{{Adibekyan} et~al.}{2012}]{adibekyan12}
{Adibekyan} V.~Z.,  {Sousa} S.~G.,  {Santos} N.~C.,  {Delgado Mena} E.,
  {Gonz{\'a}lez Hern{\'a}ndez} J.~I.,  {Israelian} G.,  {Mayor} M.,
  {Khachatryan} G.,  2012, \mn@doi [\aap] {10.1051/0004-6361/201219401}, \href
  {http://adsabs.harvard.edu/abs/2012A%26A...545A..32A} {545, A32}

\bibitem[\protect\citeauthoryear{{Adibekyan} et~al.,}{{Adibekyan}
  et~al.}{2013}]{adibekyan13}
{Adibekyan} V.~Z.,  et~al., 2013, \mn@doi [\aap] {10.1051/0004-6361/201321520},
  \href {http://adsabs.harvard.edu/abs/2013A%26A...554A..44A} {554, A44}

\bibitem[\protect\citeauthoryear{{Allende Prieto}, {Kawata}  \&
  {Cropper}}{{Allende Prieto} et~al.}{2016}]{allende-prieto16}
{Allende Prieto} C.,  {Kawata} D.,   {Cropper} M.,  2016, \mn@doi [\aap]
  {10.1051/0004-6361/201629787}, \href
  {http://adsabs.harvard.edu/abs/2016A%26A...596A..98A} {596, A98}

\bibitem[\protect\citeauthoryear{{Anders} et~al.,}{{Anders}
  et~al.}{2014}]{anders14}
{Anders} F.,  et~al., 2014, \mn@doi [\aap] {10.1051/0004-6361/201323038}, \href
  {http://adsabs.harvard.edu/abs/2014A%26A...564A.115A} {564, A115}

\bibitem[\protect\citeauthoryear{{Anders}, {Chiappini}, {Santiago},
  {Matijevi{\v c}}, {Queiroz}, {Steinmetz}  \& {Guiglion}}{{Anders}
  et~al.}{2018}]{anders18}
{Anders} F.,  {Chiappini} C.,  {Santiago} B.~X.,  {Matijevi{\v c}} G.,
  {Queiroz} A.~B.,  {Steinmetz} M.,   {Guiglion} G.,  2018, \mn@doi [\aap]
  {10.1051/0004-6361/201833099}, \href
  {http://adsabs.harvard.edu/abs/2018A%26A...619A.125A} {619, A125}

\bibitem[\protect\citeauthoryear{{Aumer} \& {White}}{{Aumer} \&
  {White}}{2013}]{aumer13a}
{Aumer} M.,  {White} S.~D.~M.,  2013, \mn@doi [\mnras] {10.1093/mnras/sts083},
  \href {http://adsabs.harvard.edu/abs/2013MNRAS.428.1055A} {428, 1055}

\bibitem[\protect\citeauthoryear{{Aumer}, {White}, {Naab}  \&
  {Scannapieco}}{{Aumer} et~al.}{2013}]{aumer13b}
{Aumer} M.,  {White} S.~D.~M.,  {Naab} T.,   {Scannapieco} C.,  2013, \mn@doi
  [\mnras] {10.1093/mnras/stt1230}, \href
  {http://adsabs.harvard.edu/abs/2013MNRAS.434.3142A} {434, 3142}

\bibitem[\protect\citeauthoryear{{Bensby} \& {Lind}}{{Bensby} \&
  {Lind}}{2018}]{bensby18}
{Bensby} T.,  {Lind} K.,  2018, \mn@doi [\aap] {10.1051/0004-6361/201833118},
  \href {http://adsabs.harvard.edu/abs/2018A%26A...615A.151B} {615, A151}

\bibitem[\protect\citeauthoryear{{Bergemann} et~al.,}{{Bergemann}
  et~al.}{2014}]{bergemann14}
{Bergemann} M.,  et~al., 2014, \mn@doi [\aap] {10.1051/0004-6361/201423456},
  \href {http://adsabs.harvard.edu/abs/2014A%26A...565A..89B} {565, A89}

\bibitem[\protect\citeauthoryear{{Binney}}{{Binney}}{2013}]{binney13}
{Binney} J.,  2013, \mn@doi [\nar] {10.1016/j.newar.2013.08.001}, \href
  {http://adsabs.harvard.edu/abs/2013NewAR..57...29B} {57, 29}

\bibitem[\protect\citeauthoryear{{Binney} \& {Tremaine}}{{Binney} \&
  {Tremaine}}{2008}]{bt08}
{Binney} J.,  {Tremaine} S.,  2008, {Galactic Dynamics: Second Edition}.
Princeton University Press

\bibitem[\protect\citeauthoryear{{Bland-Hawthorn} \&
  {Gerhard}}{{Bland-Hawthorn} \& {Gerhard}}{2016}]{bland-hawthorn16}
{Bland-Hawthorn} J.,  {Gerhard} O.,  2016, \mn@doi [\araa]
  {10.1146/annurev-astro-081915-023441}, \href
  {http://adsabs.harvard.edu/abs/2016ARA%26A..54..529B} {54, 529}

\bibitem[\protect\citeauthoryear{{Bland-Hawthorn}, {Krumholz}  \&
  {Freeman}}{{Bland-Hawthorn} et~al.}{2010}]{bland-hawthorn10}
{Bland-Hawthorn} J.,  {Krumholz} M.~R.,   {Freeman} K.,  2010, \mn@doi [\apj]
  {10.1088/0004-637X/713/1/166}, \href
  {http://adsabs.harvard.edu/abs/2010ApJ...713..166B} {713, 166}

\bibitem[\protect\citeauthoryear{{Boeche} et~al.,}{{Boeche}
  et~al.}{2013a}]{boeche13a}
{Boeche} C.,  et~al., 2013a, \mn@doi [\aap] {10.1051/0004-6361/201219607},
  \href {http://adsabs.harvard.edu/abs/2013A%26A...553A..19B} {553, A19}

\bibitem[\protect\citeauthoryear{{Boeche} et~al.,}{{Boeche}
  et~al.}{2013b}]{boeche13b}
{Boeche} C.,  et~al., 2013b, \mn@doi [\aap] {10.1051/0004-6361/201322085},
  \href {http://adsabs.harvard.edu/abs/2013A%26A...559A..59B} {559, A59}

\bibitem[\protect\citeauthoryear{{Bournaud}, {Elmegreen}  \&
  {Martig}}{{Bournaud} et~al.}{2009}]{bournaud09}
{Bournaud} F.,  {Elmegreen} B.~G.,   {Martig} M.,  2009, \mn@doi [\apjl]
  {10.1088/0004-637X/707/1/L1}, \href
  {http://adsabs.harvard.edu/abs/2009ApJ...707L...1B} {707, L1}

\bibitem[\protect\citeauthoryear{{Bovy}, {Rix}, {Schlafly}, {Nidever},
  {Holtzman}, {Shetrone}  \& {Beers}}{{Bovy} et~al.}{2016}]{bovy16}
{Bovy} J.,  {Rix} H.-W.,  {Schlafly} E.~F.,  {Nidever} D.~L.,  {Holtzman}
  J.~A.,  {Shetrone} M.,   {Beers} T.~C.,  2016, \mn@doi [\apj]
  {10.3847/0004-637X/823/1/30}, \href
  {http://cdsads.u-strasbg.fr/abs/2016ApJ...823...30B} {823, 30}

\bibitem[\protect\citeauthoryear{Brase \& Brase}{Brase \&
  Brase}{2016}]{brase16}
Brase C.,  Brase C.,  2016, Understandable Statistics: Concepts and Methods.
Cengage Learning, \url {https://books.google.de/books?id=f925DQAAQBAJ}

\bibitem[\protect\citeauthoryear{{Carney}, {Latham}  \& {Laird}}{{Carney}
  et~al.}{1990}]{carney90}
{Carney} B.~W.,  {Latham} D.~W.,   {Laird} J.~B.,  1990, \mn@doi [\aj]
  {10.1086/115351}, \href {http://adsabs.harvard.edu/abs/1990AJ.....99..572C}
  {99, 572}

\bibitem[\protect\citeauthoryear{{Carraro}, {V{\'a}zquez}, {Costa}, {Ahumada}
  \& {Giorgi}}{{Carraro} et~al.}{2015}]{carraro15}
{Carraro} G.,  {V{\'a}zquez} R.~A.,  {Costa} E.,  {Ahumada} J.~A.,   {Giorgi}
  E.~E.,  2015, \mn@doi [\aj] {10.1088/0004-6256/149/1/12}, \href
  {http://adsabs.harvard.edu/abs/2015AJ....149...12C} {149, 12}

\bibitem[\protect\citeauthoryear{{Casagrande} et~al.,}{{Casagrande}
  et~al.}{2016}]{casagrande16}
{Casagrande} L.,  et~al., 2016, \mn@doi [\mnras] {10.1093/mnras/stv2320}, \href
  {http://adsabs.harvard.edu/abs/2016MNRAS.455..987C} {455, 987}

\bibitem[\protect\citeauthoryear{{Cescutti} \& {Molaro}}{{Cescutti} \&
  {Molaro}}{2019}]{cescutti19}
{Cescutti} G.,  {Molaro} P.,  2019, \mn@doi [\mnras] {10.1093/mnras/sty2967},
  \href {http://adsabs.harvard.edu/abs/2019MNRAS.482.4372C} {482, 4372}

\bibitem[\protect\citeauthoryear{{Cheng} et~al.,}{{Cheng}
  et~al.}{2012}]{cheng12a}
{Cheng} J.~Y.,  et~al., 2012, \mn@doi [\apj] {10.1088/0004-637X/746/2/149},
  \href {http://adsabs.harvard.edu/abs/2012ApJ...746..149C} {746, 149}

\bibitem[\protect\citeauthoryear{{Chiappini}}{{Chiappini}}{2009}]{chiappini09a}
{Chiappini} C.,  2009, in {J.~Andersen, J.~Bland-Hawthorn, \& B.~Nordstr{\"o}m}
  ed.,  IAU Symposium Vol. 254, IAU Symposium. pp 191--196,
  \mn@doi{10.1017/S1743921308027580}

\bibitem[\protect\citeauthoryear{{Ciuc{\v a}}, {Kawata}, {Lin}, {Casagrande},
  {Seabroke}  \& {Cropper}}{{Ciuc{\v a}} et~al.}{2018}]{ciuca18}
{Ciuc{\v a}} I.,  {Kawata} D.,  {Lin} J.,  {Casagrande} L.,  {Seabroke} G.,
  {Cropper} M.,  2018, \mn@doi [\mnras] {10.1093/mnras/stx3285}, \href
  {http://adsabs.harvard.edu/abs/2018MNRAS.475.1203C} {475, 1203}

\bibitem[\protect\citeauthoryear{{Comer{\'o}n} et~al.,}{{Comer{\'o}n}
  et~al.}{2011}]{comeron11}
{Comer{\'o}n} S.,  et~al., 2011, \mn@doi [\apj] {10.1088/0004-637X/741/1/28},
  \href {http://adsabs.harvard.edu/abs/2011ApJ...741...28C} {741, 28}

\bibitem[\protect\citeauthoryear{{Dalton} et~al.,}{{Dalton}
  et~al.}{2012}]{dalton12}
{Dalton} G.,  et~al., 2012, in Ground-based and Airborne Instrumentation for
  Astronomy IV. p. 84460P, \mn@doi{10.1117/12.925950}

\bibitem[\protect\citeauthoryear{{Daniel} \& {Wyse}}{{Daniel} \&
  {Wyse}}{2018}]{daniel18}
{Daniel} K.~J.,  {Wyse} R.~F.~G.,  2018, \mn@doi [\mnras]
  {10.1093/mnras/sty199}, \href
  {http://adsabs.harvard.edu/abs/2018MNRAS.tmp..202D} {}

\bibitem[\protect\citeauthoryear{{De Pascale}, {Worley}, {de Laverny},
  {Recio-Blanco}, {Hill}  \& {Bijaoui}}{{De Pascale}
  et~al.}{2014}]{depascale14}
{De Pascale} M.,  {Worley} C.~C.,  {de Laverny} P.,  {Recio-Blanco} A.,  {Hill}
  V.,   {Bijaoui} A.,  2014, \mn@doi [\aap] {10.1051/0004-6361/201423767},
  \href {http://adsabs.harvard.edu/abs/2014A%26A...570A..68D} {570, A68}

\bibitem[\protect\citeauthoryear{{De Silva} et~al.,}{{De Silva}
  et~al.}{2015}]{desilva15}
{De Silva} G.~M.,  et~al., 2015, \mn@doi [\mnras] {10.1093/mnras/stv327}, \href
  {http://adsabs.harvard.edu/abs/2015MNRAS.449.2604D} {449, 2604}

\bibitem[\protect\citeauthoryear{{Delgado Mena} et~al.,}{{Delgado Mena}
  et~al.}{2015}]{delgadomena15}
{Delgado Mena} E.,  et~al., 2015, \mn@doi [\aap] {10.1051/0004-6361/201425433},
  \href {http://adsabs.harvard.edu/abs/2015A%26A...576A..69D} {576, A69}

\bibitem[\protect\citeauthoryear{{Delgado Mena}, {Tsantaki}, {Adibekyan},
  {Sousa}, {Santos}, {Gonz{\'a}lez Hern{\'a}ndez}  \& {Israelian}}{{Delgado
  Mena} et~al.}{2017}]{delgadomena17}
{Delgado Mena} E.,  {Tsantaki} M.,  {Adibekyan} V.~Z.,  {Sousa} S.~G.,
  {Santos} N.~C.,  {Gonz{\'a}lez Hern{\'a}ndez} J.~I.,   {Israelian} G.,  2017,
  \mn@doi [\aap] {10.1051/0004-6361/201730535}, \href
  {http://adsabs.harvard.edu/abs/2017A%26A...606A..94D} {606, A94}

\bibitem[\protect\citeauthoryear{{Edvardsson}, {Andersen}, {Gustafsson},
  {Lambert}, {Nissen}  \& {Tomkin}}{{Edvardsson} et~al.}{1993}]{edvardsson93}
{Edvardsson} B.,  {Andersen} J.,  {Gustafsson} B.,  {Lambert} D.~L.,  {Nissen}
  P.~E.,   {Tomkin} J.,  1993, \aap, \href
  {http://adsabs.harvard.edu/abs/1993A%26A...275..101E} {275, 101}

\bibitem[\protect\citeauthoryear{{Feast}, {Menzies}, {Matsunaga}  \&
  {Whitelock}}{{Feast} et~al.}{2014}]{feast14}
{Feast} M.~W.,  {Menzies} J.~W.,  {Matsunaga} N.,   {Whitelock} P.~A.,  2014,
  \mn@doi [\nat] {10.1038/nature13246}, \href
  {http://adsabs.harvard.edu/abs/2014Natur.509..342F} {509, 342}

\bibitem[\protect\citeauthoryear{{Forbes}, {Krumholz}  \& {Burkert}}{{Forbes}
  et~al.}{2012}]{forbes12}
{Forbes} J.,  {Krumholz} M.,   {Burkert} A.,  2012, \mn@doi [\apj]
  {10.1088/0004-637X/754/1/48}, \href
  {http://cdsads.u-strasbg.fr/abs/2012ApJ...754...48F} {754, 48}

\bibitem[\protect\citeauthoryear{{Frankel}, {Rix}, {Ting}, {Ness}  \&
  {Hogg}}{{Frankel} et~al.}{2018}]{frankel18}
{Frankel} N.,  {Rix} H.-W.,  {Ting} Y.-S.,  {Ness} M.,   {Hogg} D.~W.,  2018,
  \mn@doi [\apj] {10.3847/1538-4357/aadba5}, \href
  {http://adsabs.harvard.edu/abs/2018ApJ...865...96F} {865, 96}

\bibitem[\protect\citeauthoryear{{Freeman} \& {Bland-Hawthorn}}{{Freeman} \&
  {Bland-Hawthorn}}{2002}]{freeman02}
{Freeman} K.,  {Bland-Hawthorn} J.,  2002, \mn@doi [\araa]
  {10.1146/annurev.astro.40.060401.093840}, \href
  {http://cdsads.u-strasbg.fr/abs/2002ARA%26A..40..487F} {40, 487}

\bibitem[\protect\citeauthoryear{{Gaia Collaboration} et~al.,}{{Gaia
  Collaboration} et~al.}{2016}]{gaia16}
{Gaia Collaboration} et~al., 2016, \mn@doi [\aap]
  {10.1051/0004-6361/201629512}, \href
  {http://adsabs.harvard.edu/abs/2016A%26A...595A...2G} {595, A2}

\bibitem[\protect\citeauthoryear{{Gilmore} \& {Reid}}{{Gilmore} \&
  {Reid}}{1983}]{gilmore83}
{Gilmore} G.,  {Reid} N.,  1983, \mnras, \href
  {http://adsabs.harvard.edu/abs/1983MNRAS.202.1025G} {202, 1025}

\bibitem[\protect\citeauthoryear{{Gilmore} et~al.,}{{Gilmore}
  et~al.}{2012}]{gilmore12}
{Gilmore} G.,  et~al., 2012, The Messenger, \href
  {http://adsabs.harvard.edu/abs/2012Msngr.147...25G} {147, 25}

\bibitem[\protect\citeauthoryear{{Grand}, {Springel}, {G{\'o}mez}, {Marinacci},
  {Pakmor}, {Campbell}  \& {Jenkins}}{{Grand} et~al.}{2016}]{grand16}
{Grand} R.~J.~J.,  {Springel} V.,  {G{\'o}mez} F.~A.,  {Marinacci} F.,
  {Pakmor} R.,  {Campbell} D.~J.~R.,   {Jenkins} A.,  2016, \mn@doi [\mnras]
  {10.1093/mnras/stw601}, \href
  {http://adsabs.harvard.edu/abs/2016MNRAS.459..199G} {459, 199}

\bibitem[\protect\citeauthoryear{{Guiglion}, {de Laverny}, {Recio-Blanco},
  {Worley}, {de Pascale}, {Masseron}, {Prantzos}  \& {Mikolaitis}}{{Guiglion}
  et~al.}{2016a}]{guiglion16b}
{Guiglion} G.,  {de Laverny} P.,  {Recio-Blanco} A.,  {Worley} C.~C.,  {de
  Pascale} M.,  {Masseron} T.,  {Prantzos} N.,   {Mikolaitis} S.,  2016a,
  VizieR Online Data Catalog, \href
  {http://adsabs.harvard.edu/abs/2016yCat..35950018G} {359}

\bibitem[\protect\citeauthoryear{{Guiglion}, {de Laverny}, {Recio-Blanco},
  {Worley}, {De Pascale}, {Masseron}, {Prantzos}  \& {Mikolaitis}}{{Guiglion}
  et~al.}{2016b}]{guiglion16}
{Guiglion} G.,  {de Laverny} P.,  {Recio-Blanco} A.,  {Worley} C.~C.,  {De
  Pascale} M.,  {Masseron} T.,  {Prantzos} N.,   {Mikolaitis} {\v S}.,  2016b,
  \mn@doi [\aap] {10.1051/0004-6361/201628919}, \href
  {http://adsabs.harvard.edu/abs/2016A%26A...595A..18G} {595, A18}

\bibitem[\protect\citeauthoryear{{Guiglion} et~al.,}{{Guiglion}
  et~al.}{2019}]{guiglion19}
{Guiglion} G.,  et~al., 2019, \mn@doi [\aap] {10.1051/0004-6361/201834203},
  \href {http://adsabs.harvard.edu/abs/2019A%26A...623A..99G} {623, A99}

\bibitem[\protect\citeauthoryear{{Hayden} et~al.,}{{Hayden}
  et~al.}{2014}]{hayden14}
{Hayden} M.~R.,  et~al., 2014, \mn@doi [\aj] {10.1088/0004-6256/147/5/116},
  \href {http://adsabs.harvard.edu/abs/2014AJ....147..116H} {147, 116}

\bibitem[\protect\citeauthoryear{{Hayden}, {Recio-Blanco}, {de Laverny},
  {Mikolaitis}  \& {Worley}}{{Hayden} et~al.}{2017}]{hayden17}
{Hayden} M.~R.,  {Recio-Blanco} A.,  {de Laverny} P.,  {Mikolaitis} S.,
  {Worley} C.~C.,  2017, \mn@doi [\aap] {10.1051/0004-6361/201731494}, \href
  {http://adsabs.harvard.edu/abs/2017A%26A...608L...1H} {608, L1}

\bibitem[\protect\citeauthoryear{{Haywood}}{{Haywood}}{2008}]{haywood08}
{Haywood} M.,  2008, \mn@doi [\mnras] {10.1111/j.1365-2966.2008.13395.x}, \href
  {http://adsabs.harvard.edu/abs/2008MNRAS.388.1175H} {388, 1175}

\bibitem[\protect\citeauthoryear{{Haywood}, {Di Matteo}, {Lehnert}, {Katz}  \&
  {G{\'o}mez}}{{Haywood} et~al.}{2013}]{haywood13}
{Haywood} M.,  {Di Matteo} P.,  {Lehnert} M.~D.,  {Katz} D.,   {G{\'o}mez} A.,
  2013, \mn@doi [\aap] {10.1051/0004-6361/201321397}, \href
  {http://adsabs.harvard.edu/abs/2013A%26A...560A.109H} {560, A109}

\bibitem[\protect\citeauthoryear{{Hogg} et~al.,}{{Hogg} et~al.}{2016}]{hogg16}
{Hogg} D.~W.,  et~al., 2016, \mn@doi [\apj] {10.3847/1538-4357/833/2/262},
  \href {http://adsabs.harvard.edu/abs/2016ApJ...833..262H} {833, 262}

\bibitem[\protect\citeauthoryear{{Howell} et~al.,}{{Howell}
  et~al.}{2014}]{howell14}
{Howell} S.~B.,  et~al., 2014, \mn@doi [\pasp] {10.1086/676406}, \href
  {http://adsabs.harvard.edu/abs/2014PASP..126..398H} {126, 398}

\bibitem[\protect\citeauthoryear{{Kalberla}, {Kerp}, {Dedes}  \&
  {Haud}}{{Kalberla} et~al.}{2014}]{kalberla14}
{Kalberla} P.~M.~W.,  {Kerp} J.,  {Dedes} L.,   {Haud} U.,  2014, \mn@doi
  [\apj] {10.1088/0004-637X/794/1/90}, \href
  {http://adsabs.harvard.edu/abs/2014ApJ...794...90K} {794, 90}

\bibitem[\protect\citeauthoryear{{Kawata}, {Grand}, {Gibson}, {Casagrande},
  {Hunt}  \& {Brook}}{{Kawata} et~al.}{2017}]{kawata17}
{Kawata} D.,  {Grand} R.~J.~J.,  {Gibson} B.~K.,  {Casagrande} L.,  {Hunt}
  J.~A.~S.,   {Brook} C.~B.,  2017, \mn@doi [\mnras] {10.1093/mnras/stw2363},
  \href {http://adsabs.harvard.edu/abs/2017MNRAS.464..702K} {464, 702}

\bibitem[\protect\citeauthoryear{{Kazantzidis}, {Bullock}, {Zentner},
  {Kravtsov}  \& {Moustakas}}{{Kazantzidis} et~al.}{2008}]{kazantzidis08}
{Kazantzidis} S.,  {Bullock} J.~S.,  {Zentner} A.~R.,  {Kravtsov} A.~V.,
  {Moustakas} L.~A.,  2008, \mn@doi [\apj] {10.1086/591958}, \href
  {http://adsabs.harvard.edu/abs/2008ApJ...688..254K} {688, 254}

\bibitem[\protect\citeauthoryear{{Kollmeier} et~al.,}{{Kollmeier}
  et~al.}{2017}]{kollmeier17}
{Kollmeier} J.~A.,  et~al., 2017, preprint, \href
  {http://adsabs.harvard.edu/abs/2017arXiv171103234K} {} (\mn@eprint {arXiv}
  {1711.03234})

\bibitem[\protect\citeauthoryear{{Kordopatis} et~al.,}{{Kordopatis}
  et~al.}{2011}]{kordopatis11}
{Kordopatis} G.,  et~al., 2011, \mn@doi [\aap] {10.1051/0004-6361/201117373},
  \href {http://adsabs.harvard.edu/abs/2011A%26A...535A.107K} {535, A107}

\bibitem[\protect\citeauthoryear{{Kordopatis} et~al.,}{{Kordopatis}
  et~al.}{2013}]{kordopatis13b}
{Kordopatis} G.,  et~al., 2013, \mn@doi [\aj] {10.1088/0004-6256/146/5/134},
  \href {http://adsabs.harvard.edu/abs/2013AJ....146..134K} {146, 134}

\bibitem[\protect\citeauthoryear{{Kunder} et~al.,}{{Kunder}
  et~al.}{2017}]{kunder17}
{Kunder} A.,  et~al., 2017, \mn@doi [\aj] {10.3847/1538-3881/153/2/75}, \href
  {http://adsabs.harvard.edu/abs/2017AJ....153...75K} {153, 75}

\bibitem[\protect\citeauthoryear{{Lambert} \& {Reddy}}{{Lambert} \&
  {Reddy}}{2004}]{lambert04}
{Lambert} D.~L.,  {Reddy} B.~E.,  2004, \mn@doi [\mnras]
  {10.1111/j.1365-2966.2004.07557.x}, \href
  {http://adsabs.harvard.edu/abs/2004MNRAS.349..757L} {349, 757}

\bibitem[\protect\citeauthoryear{{Laporte}, {Johnston}, {G{\'o}mez},
  {Garavito-Camargo}  \& {Besla}}{{Laporte} et~al.}{2018}]{laporte18a}
{Laporte} C.~F.~P.,  {Johnston} K.~V.,  {G{\'o}mez} F.~A.,  {Garavito-Camargo}
  N.,   {Besla} G.,  2018, \mn@doi [\mnras] {10.1093/mnras/sty1574}, \href
  {http://adsabs.harvard.edu/abs/2018MNRAS.481..286L} {481, 286}

\bibitem[\protect\citeauthoryear{{Lee} et~al.,}{{Lee} et~al.}{2011}]{lee11}
{Lee} Y.~S.,  et~al., 2011, \mn@doi [\aj] {10.1088/0004-6256/141/3/90}, \href
  {http://adsabs.harvard.edu/abs/2011AJ....141...90L} {141, 90}

\bibitem[\protect\citeauthoryear{{Li} et~al.,}{{Li} et~al.}{2013}]{li13}
{Li} Y.-L.,  et~al., 2013, \mn@doi [\pra] {10.1103/PhysRevA.88.015804}, \href
  {http://adsabs.harvard.edu/abs/2013PhRvA..88a5804L} {88, 015804}

\bibitem[\protect\citeauthoryear{{Lin}, {Dotter}, {Ting}  \& {Asplund}}{{Lin}
  et~al.}{2018}]{lin18}
{Lin} J.,  {Dotter} A.,  {Ting} Y.-S.,   {Asplund} M.,  2018, \mn@doi [\mnras]
  {10.1093/mnras/sty709}, \href
  {http://adsabs.harvard.edu/abs/2018MNRAS.477.2966L} {477, 2966}

\bibitem[\protect\citeauthoryear{{Ma}, {Hopkins}, {Wetzel}, {Kirby},
  {Angl{\'e}s-Alc{\'a}zar}, {Faucher-Gigu{\`e}re}, {Kere{\v s}}  \&
  {Quataert}}{{Ma} et~al.}{2017}]{ma17}
{Ma} X.,  {Hopkins} P.~F.,  {Wetzel} A.~R.,  {Kirby} E.~N.,
  {Angl{\'e}s-Alc{\'a}zar} D.,  {Faucher-Gigu{\`e}re} C.-A.,  {Kere{\v s}} D.,
   {Quataert} E.,  2017, \mn@doi [\mnras] {10.1093/mnras/stx273}, \href
  {http://adsabs.harvard.edu/abs/2017MNRAS.467.2430M} {467, 2430}

\bibitem[\protect\citeauthoryear{{Mackereth} et~al.,}{{Mackereth}
  et~al.}{2017}]{mackereth17}
{Mackereth} J.~T.,  et~al., 2017, \mn@doi [\mnras] {10.1093/mnras/stx1774},
  \href {http://adsabs.harvard.edu/abs/2017MNRAS.471.3057M} {471, 3057}

\bibitem[\protect\citeauthoryear{{Majewski} et~al.,}{{Majewski}
  et~al.}{2017}]{majewski17}
{Majewski} S.~R.,  et~al., 2017, \mn@doi [\aj] {10.3847/1538-3881/aa784d},
  \href {http://adsabs.harvard.edu/abs/2017AJ....154...94M} {154, 94}

\bibitem[\protect\citeauthoryear{{Martig}, {Bournaud}, {Croton}, {Dekel}  \&
  {Teyssier}}{{Martig} et~al.}{2012}]{martig12}
{Martig} M.,  {Bournaud} F.,  {Croton} D.~J.,  {Dekel} A.,   {Teyssier} R.,
  2012, \mn@doi [\apj] {10.1088/0004-637X/756/1/26}, \href
  {http://adsabs.harvard.edu/abs/2012ApJ...756...26M} {756, 26}

\bibitem[\protect\citeauthoryear{{Martig}, {Minchev}  \& {Flynn}}{{Martig}
  et~al.}{2014}]{martig14a}
{Martig} M.,  {Minchev} I.,   {Flynn} C.,  2014, \mn@doi [\mnras]
  {10.1093/mnras/stu1003}, \href
  {http://adsabs.harvard.edu/abs/2014MNRAS.442.2474M} {442, 2474}

\bibitem[\protect\citeauthoryear{{Martig}, {Minchev}, {Ness}, {Fouesneau}  \&
  {Rix}}{{Martig} et~al.}{2016}]{martig16b}
{Martig} M.,  {Minchev} I.,  {Ness} M.,  {Fouesneau} M.,   {Rix} H.-W.,  2016,
  \mn@doi [\apj] {10.3847/0004-637X/831/2/139}, \href
  {http://adsabs.harvard.edu/abs/2016ApJ...831..139M} {831, 139}

\bibitem[\protect\citeauthoryear{{Matteucci} \& {Francois}}{{Matteucci} \&
  {Francois}}{1989}]{matteucci89}
{Matteucci} F.,  {Francois} P.,  1989, \mn@doi [\mnras]
  {10.1093/mnras/239.3.885}, \href
  {http://adsabs.harvard.edu/abs/1989MNRAS.239..885M} {239, 885}

\bibitem[\protect\citeauthoryear{{Miglio} et~al.,}{{Miglio}
  et~al.}{2017}]{miglio17}
{Miglio} A.,  et~al., 2017, \mn@doi [Astronomische Nachrichten]
  {10.1002/asna.201713385}, \href
  {http://adsabs.harvard.edu/abs/2017AN....338..644M} {338, 644}

\bibitem[\protect\citeauthoryear{{Minchev}}{{Minchev}}{2016}]{minchev16a}
{Minchev} I.,  2016, \mn@doi [Astronomische Nachrichten]
  {10.1002/asna.201612366}, \href
  {http://adsabs.harvard.edu/abs/2016AN....337..703M} {337, 703}

\bibitem[\protect\citeauthoryear{{Minchev}, {Famaey}, {Quillen}, {Dehnen},
  {Martig}  \& {Siebert}}{{Minchev} et~al.}{2012}]{minchev12b}
{Minchev} I.,  {Famaey} B.,  {Quillen} A.~C.,  {Dehnen} W.,  {Martig} M.,
  {Siebert} A.,  2012, \mn@doi [\aap] {10.1051/0004-6361/201219714}, \href
  {http://cdsads.u-strasbg.fr/abs/2012A%26A...548A.127M} {548, A127}

\bibitem[\protect\citeauthoryear{{Minchev}, {Chiappini}  \& {Martig}}{{Minchev}
  et~al.}{2013}]{mcm13}
{Minchev} I.,  {Chiappini} C.,   {Martig} M.,  2013, \mn@doi [\aap]
  {10.1051/0004-6361/201220189}, \href
  {http://cdsads.u-strasbg.fr/abs/2013A%26A...558A...9M} {558, A9}

\bibitem[\protect\citeauthoryear{{Minchev}, {Chiappini}  \& {Martig}}{{Minchev}
  et~al.}{2014a}]{mcm14}
{Minchev} I.,  {Chiappini} C.,   {Martig} M.,  2014a, \mn@doi [\aap]
  {10.1051/0004-6361/201423487}, \href
  {http://cdsads.u-strasbg.fr/abs/2014A%26A...572A..92M} {572, A92}

\bibitem[\protect\citeauthoryear{{Minchev} et~al.,}{{Minchev}
  et~al.}{2014b}]{minchev14}
{Minchev} I.,  et~al., 2014b, \apjl, \href
  {http://cdsads.u-strasbg.fr/abs/2013arXiv1310.5145M} {781, L20}

\bibitem[\protect\citeauthoryear{{Minchev}, {Martig}, {Streich}, {Scannapieco},
  {de Jong}  \& {Steinmetz}}{{Minchev} et~al.}{2015}]{minchev15}
{Minchev} I.,  {Martig} M.,  {Streich} D.,  {Scannapieco} C.,  {de Jong} R.~S.,
    {Steinmetz} M.,  2015, \mn@doi [\apjl] {10.1088/2041-8205/804/1/L9}, \href
  {http://cdsads.u-strasbg.fr/abs/2015ApJ...804L...9M} {804, L9}

\bibitem[\protect\citeauthoryear{{Minchev}, {Steinmetz}, {Chiappini}, {Martig},
  {Anders}, {Matijevic}  \& {de Jong}}{{Minchev} et~al.}{2017}]{minchev17}
{Minchev} I.,  {Steinmetz} M.,  {Chiappini} C.,  {Martig} M.,  {Anders} F.,
  {Matijevic} G.,   {de Jong} R.~S.,  2017, \mn@doi [\apj]
  {10.3847/1538-4357/834/1/27}, \href
  {http://adsabs.harvard.edu/abs/2017ApJ...834...27M} {834, 27}

\bibitem[\protect\citeauthoryear{{Minchev} et~al.,}{{Minchev}
  et~al.}{2018}]{minchev18}
{Minchev} I.,  et~al., 2018, \mn@doi [\mnras] {10.1093/mnras/sty2033}, \href
  {http://cdsads.u-strasbg.fr/abs/2018MNRAS.481.1645M} {481, 1645}

\bibitem[\protect\citeauthoryear{{Miranda} et~al.,}{{Miranda}
  et~al.}{2016}]{miranda16}
{Miranda} M.~S.,  et~al., 2016, \mn@doi [\aap] {10.1051/0004-6361/201525789},
  \href {http://adsabs.harvard.edu/abs/2016A%26A...587A..10M} {587, A10}

\bibitem[\protect\citeauthoryear{{Navarro}, {Abadi}, {Venn}, {Freeman}  \&
  {Anguiano}}{{Navarro} et~al.}{2011}]{navarro11}
{Navarro} J.~F.,  {Abadi} M.~G.,  {Venn} K.~A.,  {Freeman} K.~C.,   {Anguiano}
  B.,  2011, \mn@doi [\mnras] {10.1111/j.1365-2966.2010.17975.x}, \href
  {http://adsabs.harvard.edu/abs/2011MNRAS.412.1203N} {412, 1203}

\bibitem[\protect\citeauthoryear{{Nissen} \& {Schuster}}{{Nissen} \&
  {Schuster}}{2012}]{nissen12}
{Nissen} P.~E.,  {Schuster} W.~J.,  2012, \mn@doi [\aap]
  {10.1051/0004-6361/201219342}, \href
  {http://adsabs.harvard.edu/abs/2012A%26A...543A..28N} {543, A28}

\bibitem[\protect\citeauthoryear{{Prantzos}, {de Laverny}, {Guiglion},
  {Recio-Blanco}  \& {Worley}}{{Prantzos} et~al.}{2017}]{prantzos17}
{Prantzos} N.,  {de Laverny} P.,  {Guiglion} G.,  {Recio-Blanco} A.,   {Worley}
  C.~C.,  2017, \mn@doi [\aap] {10.1051/0004-6361/201731188}, \href
  {http://adsabs.harvard.edu/abs/2017A%26A...606A.132P} {606, A132}

\bibitem[\protect\citeauthoryear{{Queiroz} et~al.,}{{Queiroz}
  et~al.}{2018}]{queiroz18}
{Queiroz} A.~B.~A.,  et~al., 2018, \mn@doi [\mnras] {10.1093/mnras/sty330},
  \href {http://adsabs.harvard.edu/abs/2018MNRAS.tmp..326Q} {}

\bibitem[\protect\citeauthoryear{{Quinn}, {Hernquist}  \& {Fullagar}}{{Quinn}
  et~al.}{1993}]{quinn93}
{Quinn} P.~J.,  {Hernquist} L.,   {Fullagar} D.~P.,  1993, \mn@doi [\apj]
  {10.1086/172184}, \href {http://adsabs.harvard.edu/abs/1993ApJ...403...74Q}
  {403, 74}

\bibitem[\protect\citeauthoryear{{Rahimi}, {Carrell}  \& {Kawata}}{{Rahimi}
  et~al.}{2014}]{rahimi14}
{Rahimi} A.,  {Carrell} K.,   {Kawata} D.,  2014, \mn@doi [Research in
  Astronomy and Astrophysics] {10.1088/1674-4527/14/11/004}, \href
  {http://adsabs.harvard.edu/abs/2014RAA....14.1406R} {14, 1406}

\bibitem[\protect\citeauthoryear{{Ram{\'{\i}}rez}, {Fish}, {Lambert}  \&
  {Allende Prieto}}{{Ram{\'{\i}}rez} et~al.}{2012}]{ramirez12}
{Ram{\'{\i}}rez} I.,  {Fish} J.~R.,  {Lambert} D.~L.,   {Allende Prieto} C.,
  2012, \mn@doi [\apj] {10.1088/0004-637X/756/1/46}, \href
  {http://adsabs.harvard.edu/abs/2012ApJ...756...46R} {756, 46}

\bibitem[\protect\citeauthoryear{{Rauer} et~al.,}{{Rauer}
  et~al.}{2014}]{rauer14}
{Rauer} H.,  et~al., 2014, \mn@doi [Experimental Astronomy]
  {10.1007/s10686-014-9383-4}, \href
  {http://adsabs.harvard.edu/abs/2014ExA....38..249R} {38, 249}

\bibitem[\protect\citeauthoryear{{Rebolo}, {Molaro}  \& {Beckman}}{{Rebolo}
  et~al.}{1988}]{rebolo88}
{Rebolo} R.,  {Molaro} P.,   {Beckman} J.~E.,  1988, \aap, \href
  {http://adsabs.harvard.edu/abs/1988A%26A...192..192R} {192, 192}

\bibitem[\protect\citeauthoryear{{Recio-Blanco} et~al.,}{{Recio-Blanco}
  et~al.}{2014}]{recio-blanco14}
{Recio-Blanco} A.,  et~al., 2014, \mn@doi [\aap] {10.1051/0004-6361/201322944},
  \href {http://adsabs.harvard.edu/abs/2014A%26A...567A...5R} {567, A5}

\bibitem[\protect\citeauthoryear{{Ricker} et~al.,}{{Ricker}
  et~al.}{2015}]{ricker15}
{Ricker} G.~R.,  et~al., 2015, \mn@doi [Journal of Astronomical Telescopes,
  Instruments, and Systems] {10.1117/1.JATIS.1.1.014003}, \href
  {http://adsabs.harvard.edu/abs/2015JATIS...1a4003R} {1, 014003}

\bibitem[\protect\citeauthoryear{{Rix} \& {Bovy}}{{Rix} \&
  {Bovy}}{2013}]{rix13}
{Rix} H.-W.,  {Bovy} J.,  2013, \mn@doi [\aapr] {10.1007/s00159-013-0061-8},
  \href {http://adsabs.harvard.edu/abs/2013A%26ARv..21...61R} {21, 61}

\bibitem[\protect\citeauthoryear{{Rocha-Pinto}, {Rangel}, {Porto de Mello},
  {Bragan{\c c}a}  \& {Maciel}}{{Rocha-Pinto} et~al.}{2006}]{rocha-pinto06}
{Rocha-Pinto} H.~J.,  {Rangel} R.~H.~O.,  {Porto de Mello} G.~F.,  {Bragan{\c
  c}a} G.~A.,   {Maciel} W.~J.,  2006, \mn@doi [\aap]
  {10.1051/0004-6361:20065362}, \href
  {http://adsabs.harvard.edu/abs/2006A%26A...453L...9R} {453, L9}

\bibitem[\protect\citeauthoryear{{Romano}, {Matteucci}, {Molaro}  \&
  {Bonifacio}}{{Romano} et~al.}{1999}]{romano99}
{Romano} D.,  {Matteucci} F.,  {Molaro} P.,   {Bonifacio} P.,  1999, \aap,
  \href {http://adsabs.harvard.edu/abs/1999A%26A...352..117R} {352, 117}

\bibitem[\protect\citeauthoryear{{Ro{\v s}kar}, {Debattista}, {Quinn},
  {Stinson}  \& {Wadsley}}{{Ro{\v s}kar} et~al.}{2008}]{roskar08a}
{Ro{\v s}kar} R.,  {Debattista} V.~P.,  {Quinn} T.~R.,  {Stinson} G.~S.,
  {Wadsley} J.,  2008, \mn@doi [\apjl] {10.1086/592231}, \href
  {http://adsabs.harvard.edu/abs/2008ApJ...684L..79R} {684, L79}

\bibitem[\protect\citeauthoryear{{Ro{\v s}kar}, {Debattista}, {Brooks},
  {Quinn}, {Brook}, {Governato}, {Dalcanton}  \& {Wadsley}}{{Ro{\v s}kar}
  et~al.}{2010}]{roskar10}
{Ro{\v s}kar} R.,  {Debattista} V.~P.,  {Brooks} A.~M.,  {Quinn} T.~R.,
  {Brook} C.~B.,  {Governato} F.,  {Dalcanton} J.~J.,   {Wadsley} J.,  2010,
  \mn@doi [\mnras] {10.1111/j.1365-2966.2010.17178.x}, \href
  {http://adsabs.harvard.edu/abs/2010MNRAS.408..783R} {408, 783}

\bibitem[\protect\citeauthoryear{{Ro{\v s}kar}, {Debattista}  \&
  {Loebman}}{{Ro{\v s}kar} et~al.}{2013}]{roskar13}
{Ro{\v s}kar} R.,  {Debattista} V.~P.,   {Loebman} S.~R.,  2013, \mn@doi
  [\mnras] {10.1093/mnras/stt788}, \href
  {http://adsabs.harvard.edu/abs/2013MNRAS.433..976R} {433, 976}

\bibitem[\protect\citeauthoryear{{Santiago} et~al.,}{{Santiago}
  et~al.}{2016}]{santiago16}
{Santiago} B.~X.,  et~al., 2016, \mn@doi [\aap] {10.1051/0004-6361/201323177},
  \href {http://adsabs.harvard.edu/abs/2016A%26A...585A..42S} {585, A42}

\bibitem[\protect\citeauthoryear{{Scannapieco}, {White}, {Springel}  \&
  {Tissera}}{{Scannapieco} et~al.}{2009}]{scannapieco09}
{Scannapieco} C.,  {White} S.~D.~M.,  {Springel} V.,   {Tissera} P.~B.,  2009,
  \mn@doi [\mnras] {10.1111/j.1365-2966.2009.14764.x}, \href
  {http://adsabs.harvard.edu/abs/2009MNRAS.396..696S} {396, 696}

\bibitem[\protect\citeauthoryear{{Schlesinger} et~al.,}{{Schlesinger}
  et~al.}{2011}]{schlesinger12}
{Schlesinger} K.~J.,  et~al., 2011, arXiv:1112.2214, \href
  {http://adsabs.harvard.edu/abs/2011arXiv1112.2214S} {}

\bibitem[\protect\citeauthoryear{{Schlesinger} et~al.,}{{Schlesinger}
  et~al.}{2014}]{schlesinger14}
{Schlesinger} K.~J.,  et~al., 2014, \mn@doi [\apj]
  {10.1088/0004-637X/791/2/112}, \href
  {http://adsabs.harvard.edu/abs/2014ApJ...791..112S} {791, 112}

\bibitem[\protect\citeauthoryear{{Sch{\"o}nrich} \& {Binney}}{{Sch{\"o}nrich}
  \& {Binney}}{2009a}]{schonrich09a}
{Sch{\"o}nrich} R.,  {Binney} J.,  2009a, \mn@doi [\mnras]
  {10.1111/j.1365-2966.2009.14750.x}, \href
  {http://adsabs.harvard.edu/abs/2009MNRAS.396..203S} {396, 203}

\bibitem[\protect\citeauthoryear{{Sch{\"o}nrich} \& {Binney}}{{Sch{\"o}nrich}
  \& {Binney}}{2009b}]{schonrich09b}
{Sch{\"o}nrich} R.,  {Binney} J.,  2009b, \mn@doi [\mnras]
  {10.1111/j.1365-2966.2009.15365.x}, \href
  {http://adsabs.harvard.edu/abs/2009MNRAS.399.1145S} {399, 1145}

\bibitem[\protect\citeauthoryear{{Sch{\"o}nrich} \& {McMillan}}{{Sch{\"o}nrich}
  \& {McMillan}}{2017}]{schonrich17}
{Sch{\"o}nrich} R.,  {McMillan} P.~J.,  2017, \mn@doi [\mnras]
  {10.1093/mnras/stx093}, \href
  {http://adsabs.harvard.edu/abs/2017MNRAS.467.1154S} {467, 1154}

\bibitem[\protect\citeauthoryear{{Sellwood} \& {Binney}}{{Sellwood} \&
  {Binney}}{2002}]{sellwood02}
{Sellwood} J.~A.,  {Binney} J.~J.,  2002, \mn@doi [\mnras]
  {10.1046/j.1365-8711.2002.05806.x}, \href
  {http://adsabs.harvard.edu/abs/2002MNRAS.336..785S} {336, 785}

\bibitem[\protect\citeauthoryear{{Selvitella}}{{Selvitella}}{2017}]{selvitella17}
{Selvitella} A.,  2017, \mn@doi [Journal of Mathematical Physics]
  {10.1063/1.4977784}, \href
  {http://adsabs.harvard.edu/abs/2017JMP....58c2101S} {58, 032101}

\bibitem[\protect\citeauthoryear{Simpson}{Simpson}{1951}]{simpson51}
Simpson E.~H.,  1951, Journal of the Royal Statistical Society. Series B
  (Methodological), 13, 238

\bibitem[\protect\citeauthoryear{Smith \& Goltz}{Smith \&
  Goltz}{2012}]{smith12}
Smith M.~L.,  Goltz H.~H.,  2012, \mn@doi [Health Promotion Practice]
  {10.1177/1524839911419290}, 13, 637

\bibitem[\protect\citeauthoryear{{Spagna}, {Lattanzi}, {Re Fiorentin}  \&
  {Smart}}{{Spagna} et~al.}{2010}]{spagna10}
{Spagna} A.,  {Lattanzi} M.~G.,  {Re Fiorentin} P.,   {Smart} R.~L.,  2010,
  \mn@doi [\aap] {10.1051/0004-6361/200913538}, \href
  {http://adsabs.harvard.edu/abs/2010A%26A...510L...4S} {510, L4}

\bibitem[\protect\citeauthoryear{{Spitoni}, {Cescutti}, {Minchev}, {Matteucci},
  {Silva Aguirre}, {Martig}, {Bono}  \& {Chiappini}}{{Spitoni}
  et~al.}{2018}]{spitoni19}
{Spitoni} E.,  {Cescutti} G.,  {Minchev} I.,  {Matteucci} F.,  {Silva Aguirre}
  V.,  {Martig} M.,  {Bono} G.,   {Chiappini} C.,  2018, arXiv e-prints, \href
  {http://adsabs.harvard.edu/abs/2018arXiv181111196S} {}

\bibitem[\protect\citeauthoryear{{Springel} et~al.,}{{Springel}
  et~al.}{2008}]{springel08}
{Springel} V.,  et~al., 2008, \mn@doi [\mnras]
  {10.1111/j.1365-2966.2008.14066.x}, \href
  {http://adsabs.harvard.edu/abs/2008MNRAS.391.1685S} {391, 1685}

\bibitem[\protect\citeauthoryear{{Thomas} et~al.,}{{Thomas}
  et~al.}{2019}]{thomas19}
{Thomas} G.~F.,  et~al., 2019, \mn@doi [\mnras] {10.1093/mnras/sty3334}, \href
  {http://adsabs.harvard.edu/abs/2019MNRAS.483.3119T} {483, 3119}

\bibitem[\protect\citeauthoryear{{Thompson}}{{Thompson}}{2006}]{thompson06}
{Thompson} B.,  2006, Foundations of behavioral statistics: An insight-based
  approach. New York: Guilford, \href
  {https://psycnet.apa.org/record/2006-07704-000} {}

\bibitem[\protect\citeauthoryear{{Vera-Ciro}, {D'Onghia}, {Navarro}  \&
  {Abadi}}{{Vera-Ciro} et~al.}{2014}]{vera-ciro14}
{Vera-Ciro} C.,  {D'Onghia} E.,  {Navarro} J.,   {Abadi} M.,  2014, \mn@doi
  [\apj] {10.1088/0004-637X/794/2/173}, \href
  {http://adsabs.harvard.edu/abs/2014ApJ...794..173V} {794, 173}

\bibitem[\protect\citeauthoryear{{Vickers} \& {Smith}}{{Vickers} \&
  {Smith}}{2018}]{vickers18}
{Vickers} J.~J.,  {Smith} M.~C.,  2018, \mn@doi [\apj]
  {10.3847/1538-4357/aac323}, \href
  {http://adsabs.harvard.edu/abs/2018ApJ...860...91V} {860, 91}

\bibitem[\protect\citeauthoryear{{Villalobos} \& {Helmi}}{{Villalobos} \&
  {Helmi}}{2008}]{villalobos08}
{Villalobos} {\'A}.,  {Helmi} A.,  2008, \mn@doi [\mnras]
  {10.1111/j.1365-2966.2008.13979.x}, \href
  {http://adsabs.harvard.edu/abs/2008MNRAS.391.1806V} {391, 1806}

\bibitem[\protect\citeauthoryear{{Wan}, {Liu}  \& {Deng}}{{Wan}
  et~al.}{2017}]{wan17}
{Wan} J.-C.,  {Liu} C.,   {Deng} L.-C.,  2017, \mn@doi [Research in Astronomy
  and Astrophysics] {10.1088/1674-4527/17/8/79}, \href
  {http://adsabs.harvard.edu/abs/2017RAA....17...79W} {17, 079}

\bibitem[\protect\citeauthoryear{{Wang}, {Liu}, {Xu}, {Wan}  \& {Deng}}{{Wang}
  et~al.}{2018}]{wang18}
{Wang} H.-F.,  {Liu} C.,  {Xu} Y.,  {Wan} J.-C.,   {Deng} L.,  2018, \mn@doi
  [\mnras] {10.1093/mnras/sty1058}, \href
  {http://adsabs.harvard.edu/abs/2018MNRAS.478.3367W} {478, 3367}

\bibitem[\protect\citeauthoryear{{Wang} et~al.,}{{Wang} et~al.}{2019}]{wang19}
{Wang} C.,  et~al., 2019, \mn@doi [\mnras] {10.1093/mnras/sty2797}, \href
  {http://adsabs.harvard.edu/abs/2019MNRAS.482.2189W} {482, 2189}

\bibitem[\protect\citeauthoryear{{Xiang} et~al.,}{{Xiang}
  et~al.}{2018}]{xiang18}
{Xiang} M.,  et~al., 2018, \mn@doi [\apjs] {10.3847/1538-4365/aad237}, \href
  {http://adsabs.harvard.edu/abs/2018ApJS..237...33X} {237, 33}

\bibitem[\protect\citeauthoryear{Yule}{Yule}{1902}]{yule02}
Yule G.~U.,  1902, Biometrika, 121, 134

\bibitem[\protect\citeauthoryear{{de Grijs}}{{de Grijs}}{1998}]{degrijs98}
{de Grijs} R.,  1998, \mn@doi [\mnras] {10.1046/j.1365-8711.1998.01896.x},
  \href {http://adsabs.harvard.edu/abs/1998MNRAS.299..595D} {299, 595}

\bibitem[\protect\citeauthoryear{{de Jong} et~al.,}{{de Jong}
  et~al.}{2012}]{dejong12}
{de Jong} R.~S.,  et~al., 2012, \mn@doi [\procspie] {10.1117/12.926239}, \href
  {http://adsabs.harvard.edu/abs/2012SPIE.8446E..0TD} {8446, 84460T}

\bibitem[\protect\citeauthoryear{{de Laverny}, {Recio-Blanco}, {Worley}, {De
  Pascale}, {Hill}  \& {Bijaoui}}{{de Laverny} et~al.}{2013}]{delaverny13}
{de Laverny} P.,  {Recio-Blanco} A.,  {Worley} C.~C.,  {De Pascale} M.,  {Hill}
  V.,   {Bijaoui} A.,  2013, The Messenger, \href
  {http://adsabs.harvard.edu/abs/2013Msngr.153...18D} {153, 18}

\bibitem[\protect\citeauthoryear{{van der Kruit} \& {Searle}}{{van der Kruit}
  \& {Searle}}{1982}]{vanderkruit82}
{van der Kruit} P.~C.,  {Searle} L.,  1982, \aap, \href
  {http://adsabs.harvard.edu/abs/1982A%26A...110...61V} {110, 61}

\makeatother
\end{thebibliography}

\end{document}